\documentclass[12pt,a4paper]{article}
\pdfoutput=1
\usepackage[utf8]{inputenc}
\usepackage[T1]{fontenc}
\usepackage{style}
\usepackage[english]{babel}
\usepackage{url}
\usepackage{footnote}
\usepackage{float}
\usepackage{multirow}

\usepackage{amsmath}
\usepackage{graphicx} 
\usepackage{verbatim} 

\newcommand{\dg}{\text{deg}}

\newcommand{\kms}{\rm \,\,km\,s^{-1}Mpc^{-1}}
\newcommand{\logg}{{\rm Log_{10}}}

\renewcommand{\l}{\lambda}
\renewcommand{\d}{\partial}

\newcommand{\rdrag}{r_{\rm drag}}

\newcommand{\be}{\begin{equation}}
\newcommand{\ee}{\end{equation}}
\newcommand{\beqa}{\begin{eqnarray}}
\newcommand{\eeqa}{\end{eqnarray}}
\newcommand{\bsm}{\begin{smallmatrix}}
\newcommand{\esm}{\end{smallmatrix}}

\renewcommand\r{\rho}

\renewcommand\l{\lambda}

\def\d{\partial}
\newcommand{\bseq}{\begin{subequations}}
\newcommand{\eseq}{\end{subequations}}

\renewcommand{\ln}{\mathop{\rm ln}\nolimits}

\newcommand{\eV}{\,\text{eV}}

\renewcommand{\d}{\partial}

\def\l{\left(}
\def\r{\right)}

\usepackage[textwidth=3cm,textsize=footnotesize]{todonotes}

\title{
Combined analysis of Planck and SPTPol data favors the early dark
energy models
}

\author[a,b]{Anton Chudaykin,\footnote{\texttt{chudy@ms2.inr.ac.ru}}}
\author[a,b]{Dmitry Gorbunov,\footnote{\texttt{gorby@ms2.inr.ac.ru}}}
\author[a,c]{Nikita Nedelko\footnote{\texttt{nikita.nedelko1999@gmail.com}}}
\affiliation[a]{Institute for Nuclear Research of the
Russian Academy of Sciences, \\ 
\normalsize \it  60th October Anniversary Prospect, 7a, 117312
Moscow, Russia}
\affiliation[b]{Moscow Institute of Physics and Technology,\\
    Institutsky lane 9, Dolgoprudny, Moscow region, 141700, Russia}
\affiliation[c]{Department of Particle Physics and Cosmology, Physics Faculty,\\
    M.V. Lomonosov Moscow State University,\\
    Vorobjevy Gory, 119991 Moscow, Russia}

\abstract{We study the implications of the Planck temperature power spectrum at low multipoles, $\ell<1000$, and SPTPol data. We show that this combination predicts consistent lensing-induced smoothing of acoustic peaks within $\Lambda$CDM cosmology and yields the robust predictions of the cosmological parameters. Combining only the Planck large-scale temperature data and the
  SPTPol polarization and lensing measurements within $\Lambda$CDM
  model we found substantially lower values of linear matter density perturbation $\sigma_8$ which bring the late-time parameter
  $S_8=\sigma_8\sqrt{\Omega_m/0.3}=0.763\pm0.022$ into accordance with galaxy clustering and weak lensing measurements. It also raises up the
  Hubble constant $H_0=69.68\pm1.00\kms$ that reduces the Hubble tension
  to the $2.5\sigma$ level. We examine the residual tension in the
  Early Dark Energy (EDE) model which produces the brief energy
  injection prior to recombination. We implement both the background
  and perturbation evolutions of the scalar field which potential
  scales as $V(\phi)\propto \phi^{2n}$. Including cosmic shear measurements (KiDS, VIKING-450, DES) and local distance-ladder data (SH0ES) to the combined fit we found that EDE completely alleviates the Hubble tension while not degradating the fit to large-scale structure data. The EDE scenario significantly improves the
  goodness-of-fit by $2.9\sigma$ in comparison with the concordance
  $\Lambda$CDM model. The account for the intermediate-redshift data
  (the supernova dataset and baryon acoustic oscillation data) fits
  perfectly to our parameter predictions and indicates the preference
  of EDE over $\Lambda$CDM at $3\sigma$.}

\begin{document}

\begin{flushright}
    INR-TH-2020-019
\end{flushright}

\maketitle
\flushbottom

\section{Introduction}
\label{sec:intro}

Measurements of the cosmic microwave background (CMB) have provided a
profound insight into the nature of the Universe. Temperature and
polarization anisotropies of the CMB encode detailed information about the
composition and evolution of our Universe. Indeed, the primary CMB
is one of the most powerful probes of the early-Universe cosmology coming
from the last scattering. It also provides cosmological (model-dependent)
probe of the late-Universe composition through observation of the
angular scale of the sound horizon at last scattering
$\theta_s$. Besides, the late-time physics leaves distinctive imprints on
the CMB anisotropy at small angular scales. In particular, the small
scale distortions caused by gravitational lensing \cite{Lewis:2006fu} and
Sunyaev--Zeldovich effect \cite{Sunyaev:1970eu} provide a unique probe of the Universe
evolution at late times. Our capability in tackling all these subtle
effects strongly depends on the quality and self-consistency of
observed CMB maps.

The most profound observation of CMB anisotropy over the full sky has
been accomplished by the Planck satellite \cite{Aghanim:2018eyx}. This
mission provides with measurements of cosmological parameters at the
percent level accuracy thus exemplifying the potential of CMB surveys
as a high-precision probe of cosmology. However, the extensive
analyses of the plethora of CMB data from this spacecraft have
revealed internal tensions within the Planck dataset. The most
prominent feature often called 'lensing tension' refers to overly
enhanced lensing smoothing of the CMB peaks compared to the
$\Lambda$CDM expectation
\cite{Ade:2015xua,Addison:2015wyg,Aghanim:2016sns,Aghanim:2018eyx}. A
recently developed technique for probing the gravitational lensing
potential from CMB data in a model-independent way reveals $2.8\sigma$
tension between the full Planck dataset and $\Lambda$CDM expectation
\cite{Motloch:2019gux}. This anomaly raises the issue of whether the
Planck dataset is internally consistent. The Planck collaboration has
thoroughly examined this problem and presented the cosmological
parameter constraints from different multipole ranges
\cite{Aghanim:2016sns,Aghanim:2018eyx}. They found a moderate
disagreement ($\gtrsim 2\sigma$) in parameters extracted from
multipoles $\ell<800$ and $\ell>800$ of CMB temperature anisotropy
power spectrum (TT). The similar trend in driving cosmological parameters has been
revealed for the full Planck dataset including polarization
measurements albeit with a lower statistical significance ($\lesssim
2\sigma$). 
Although the Planck collaboration argued that the exclusion of power
spectrum at high multipoles does not change their baseline result, its
impact might be crucial for extensions of $\Lambda$CDM model. A
considerably less tension originates from the overall deficit of the
Planck TT power spectra across the multipole range
$20\lesssim\ell\lesssim30$. Remarkably, this feature is strongly
disfavoured by extra smoothing of acoustic CMB peaks observed by
Planck on small angular scales which pull the amplitude of power
spectra to larger values. These internal tensions within the Planck
dataset emphasize the need for independent measurement of the CMB
anisotropies, especially on small scales.

An important consistency check of CMB anisotropy measurements at small scales can be provided by ground-based
telescopes. The small-scale CMB anisotropies can be measured by the South Pole Telescope (SPT) \cite{Lueker:2009rx,Shirokoff:2010cs,Reichardt:2011yv} and the Atacama Cosmology Telescope (ACT) \cite{Fowler:2010cy,Das:2010ga}. The small-scale EE power spectrum has been
recently measured by POLARBEAR instrument \cite{Adachi:2020knf}. These observations supplement the satellite-based measurements since the ground-based telescopes are sensitive to much
smaller angular scales unattainable in full sky surveys. The most
sensitive measurements to date of small-angular scale temperature
anisotropy have been performed by the SPT
observation of 2540 $\dg^2$ SPT-SZ survey \cite{Story:2012wx}. The SPT
measurements of CMB temperature anisotropy at multipole range
$650<\ell<3000$ augmented with WMAP-7 observations of large angular
scales $\ell\lesssim500$ prefer a slightly smaller lensing amplitude
which is at the lower end, but within the $1\sigma$ prediction of
$\Lambda$CDM model \cite{Story:2012wx}. Thereby, SPT-SZ survey
provides consistent temperature power spectrum down to arcminute
scales which is compatible with theoretical predictions of the
$\Lambda$CDM model.

Measurements of CMB polarization anisotropies is a promising tool to obtain more robust parameter inference. Since E-mode measurements are
expected to be fractionally less contaminated by foregrounds than
temperature measurements \cite{Seiffert:2006vh,Battye:2010zz},  
the E-mode auto-power spectrum (EE) and the
temperature-E-mode correlation (TE) have higher potential to study
very small angular scales. In particular, polarization measurements
demonstrates high sensitivity to the photon diffusion damping tail of
the CMB power spectrum enabling tighter constraints on cosmological
parameters. Besides, TE and EE power spectra provide a powerful
consistency check of lensing smoothing effect which can shed light on
the internal tensions in the Planck dataset.

The most accurate measurements to date of TE and EE spectra have been
provided by SPTPol analysis of 500 $\dg^2$ survey
\cite{Henning:2017nuy}. These data are the most sensitive measurements
of TE and EE spectra at $\ell>1050$ and $\ell>1475$,
respectively. These observations reveal less smoothing of CMB acoustic
peaks as allowed by $\Lambda$CDM which level is $1.4\sigma$ below the
$\Lambda$CDM expectation and $2.9\sigma$ lower than the value prefered
by the Planck collaboration. It disproves stronger smoothing of the
acoustic scales observed in the Planck maps. Altogether, the
most sophisticated ground-based experiments provide consistent
measurements of the temperature and polarization CMB anisotropies which
do not find compelling evidence for the enhanced smoothing of the
acoustic peaks. Conversely, they found a mild deficit of the lensing
power in the CMB anisotropies on small scales which may be hints of
new physics.


Besides measurements of the CMB power spectrum, the lensing potential can be
directly extracted from quadratic estimators of $T$-, $E$- or
$B$-fields.
Independent measurement of lensing power spectrum $C_\ell^{\phi\phi}$ represents an
important cross-check of the CMB anisotropies on small scales and can
verify the lensing tension observed in the Planck maps. Lensing potential
power spectrum $C_\ell^{\phi\phi}$ in the multipole range
$100<\ell<2000$ was recently constrained from the 500$\deg$ SPTPol
survey \cite{Wu:2019hek,Bianchini:2019vxp}. This measurement based on
a minimum-variance estimator that combines both temperature and
polarization CMB maps predicts a $1.8\sigma$ lower value of lensing
power as compared to the Planck $\Lambda$CDM cosmology. Thereby, SPTPol
measurements exhibit similar trends in both CMB power spectra and
lensing potential power spectrum probes. Planck collaboration
also provides independent measurements of lensing potential power spectrum which
is consistent within $1\sigma$ to the $\Lambda$CDM expectations \cite{Aghanim:2018eyx}. Although
Planck lensing likelihood has somewhat higher constraining power \cite{Bianchini:2019vxp}, 
we exploit SPTPol lensing measurements in our analysis to extract lensing information entirely from SPTPol
survey. It allows us to investigate the deficit of lensing power observed in both TE, EE power spectra and lensing power spectrum $C_\ell^{\phi\phi}$ within SPTPol survey in a more consistent way.

Alongside with the CMB measurements which bring only model-dependent
parameter constraints in the late Universe, the local measurements
support direct extraction of cosmological parameters at low
redshifts. These probes also provide another important consistency
check of CMB results in a concrete cosmological
model. Intriguingly, there is a modest level of discrepancy between
the Planck results and direct probes within the $\Lambda$CDM model. In
particular, the amplitude of linear density fluctuations $\sigma_8$
deduced from galaxy cluster counts
\cite{Bocquet:2018ukq,Ade:2015fva,Vikhlinin:2008ym,Bohringer:2014ooa,Boehringer:2017wvr},
cosmic shear and/or galaxy clustering measurements
\cite{Hildebrandt:2016iqg,Hildebrandt:2018yau,Abbott:2017wau,Abbott:2018ydy,Hikage:2018qbn}
is distinctly lower than the value of $\sigma_8$ suggested by
Planck. Combining different cosmic shear surveys (KiDS, VIKING-450 and
DES) \cite{Joudaki:2019pmv} \footnote{Actually, the DES redshifts in
  the joint analysis \cite{Joudaki:2019pmv} was recalibrated using
  deep public spectroscopic surveys. Adopting these revised redshifts,
  the results give a $0.8\sigma$ reduction in the DES-inferred value
  of $S_8$ making the DES results compatible with KiDS and VIKING
  \cite{Hildebrandt:2016iqg,Hildebrandt:2018yau}.} leads to the
substantially tighter constraint $S_8=0.762_{-0.024}^{+0.025}$ where
$S_8=\sigma_8\sqrt{\Omega_m/0.3}$ is the principal-component parameter
for weak gravitational lensing analyses. This measurement is in
tension with the Planck value $S_8=0.830\pm0.013$ \cite{Aghanim:2018eyx} at the level of $2.5\sigma$. Recently,
this joint analysis has been improved by including additional
small-scale information yielding $S_8=0.755_{-0.021}^{+0.019}$, that
exacerbates the tension with the Planck to $3.2\sigma$
\cite{Asgari:2019fkq}. In our work, 
we stick to the joint cosmic shear analysis \cite{Joudaki:2019pmv} which applies a more conservative approach of excluding the small-scale information. 

More controversial tension between low-redshift and high-redshift data
is attributed to the determination of the present-day expansion rate of
the Universe. Planck measurements based on CMB temperature, polarization and lensing power spectra accommodate the model-dependent estimate $H_0=67.36\pm0.54$ \cite{Aghanim:2018eyx}. Traditional distance-ladder measurements made up
Cepheids and Type Ia Supernovae favour generally higher values of the
Hubble constant
\cite{Freedman:2000cf,Freedman:2012ny,Riess:2016jrr}. Upon improved 
calibration of the Cepheid distance-ladder in the Large Magellanic
Cloud, the SH0ES collaboration presents the most severe constraint $H_0=74.03\pm1.42\kms$ which
tightens the tension with the Planck measurement to $4.4\sigma$
\cite{Riess:2019cxk}. Remarkably, the direct measurement based on Type
Ia Supernovae is quite robust against the choice of distance
indicators. A number of other techniques which has been used to calibrate the SN luminosity distances (distant megamasers \cite{Pesce:2020xfe}, Miras or variable red giant stars \cite{Huang:2019yhh}, strongly lensed quasars \cite{Taubenberger:2019qna}) gives coherent results competitive with the Cepheid-based measurement \cite{Riess:2019cxk} albeit with lager $H_0$ uncertainties, and only the Tip of the Red Giant Branch (TRGB) as a distance measure brings moderately smaller value of the Hubble constant $H_0=69.6\pm1.9\kms$ \cite{Freedman:2020dne}. Intriguingly,
all distance-ladder measurements \footnote{The recent TRGB result \cite{Freedman:2020dne} exhibit somewhat lower agreement with the analysis of gravitationally lensed quasars at the level of $1.4\sigma$.} are perfectly consistent with a completely different technique of the Hubble constant measurement based on strong
gravitational lensing time delays which dictates
$H_0=73.3^{+1.7}_{-1.8}\kms$ \cite{Wong:2019kwg}. A cosmology model-independent 
analysis for a flat Universe accumulating strong gravitation lensing
and supernova data supports this estimate yielding
$H_0=72.8^{+1.6}_{-1.7}\kms$ \cite{Liao:2020zko}. 

Besides direct measurements based on local distance anchors, there is a powerful inverse-distance-ladder approach which makes use of the baryon acoustic oscillations (BAO) measurements and an independent determination of the sound horizon $\rdrag$. For the base-$\Lambda$CDM model, this approach can be used to constrain $H_0$ without using any CMB measurements. The combination of BAO data and primordial deuterium abundance measurements augmented with supernova distance data \cite{Lemos:2018smw} or the late-time probe of the matter density \cite{Abbott:2017smn} leads to the Hubble constraints, in perfect agreement with the Planck measurements. The anisotropic BAO measurements come from galaxy clustering and the Ly$\alpha$ forest along with a precise estimate of the primordial deuterium abundance yield nearly the same Hubble constant \cite{Schoneberg:2019wmt,Cuceu:2019for,Addison:2017fdm}. It is worth nothing that the recent full-shape power spectrum analyses \cite{Ivanov:2019pdj,DAmico:2019fhj,Troster:2019ean} also bring the Hubble measurement into agreement with the Planck data. However, all these measurements are obtained in the context of concordance $\Lambda$CDM cosmological model and hence they represent highly model-dependent estimates of $H_0$. Given this reason, in our work we focus on the local Cepheid-based distance-ladder measurement \cite{Riess:2019cxk} which does not rely on early universe physics and being the most widely used local probe to date. 

In turn, the ground-based measurements of the CMB anisotropies dictate
the estimates consistent with local measurements. Comparison between
the Planck and SPT-SZ temperature power spectra reveals that adding
the small-scale data from SPT-SZ survey patch drives $2\sigma$ shifts
in various cosmological parameters from the Planck expectation values
\cite{Aylor:2017haa}. In particular, the temperature power spectrum
with WMAP7-based Gaussian prior on $\tau$ yields $H_0=75.5\pm 3.5\kms$
and $\sigma_8=0.772\pm 0.035$ \cite{Story:2012wx}. While the amplitude
of density fluctuation $\sigma_8$ agrees well with both the Planck
results and the direct probes at low redshifts, the $H_0$ measurement
exhibits more than $2\sigma$ discrepancy with the Planck value being
fully consistent with the distance-ladder estimate
\cite{Riess:2019cxk}. Measurements of CMB polarization power spectra exhibit a
similar trend for the $\Lambda$CDM parameters to drift away from the
Planck values as the SPTPol range is extended to higher
multipoles. The SPTPol baseline result reads $H_0=71.29\pm 2.12\kms$
and $\sigma_8=0.771\pm 0.024$ \cite{Henning:2017nuy} that reveals a
mild preference for lower $\sigma_8$ and a $1.8\sigma$ upward shift in
$H_0$ from the Planck value.

Above findings demonstrate a perfect consistency of the CMB
measurements from the ground with the local probes at low
redshifts \footnote{The similar concordance between Planck large-scale temperature anisotropies and direct measurements of linear matter density perturbation was obtained in Ref. \cite{Burenin:2018nuf}. They claim $3.7\sigma$ tension between the Planck CMB temperature power spectrum at high multipoles $\ell>1000$ and various probes of clustering statistics.}. On the other hand, the parameter fit based on the
ground-based telescopes reveal a distinctive difference with the
Planck baseline cosmology. The bulk of this discrepancy is caused by
an excess of the lensing-induce smoothing in the Planck maps at small
scales. Since the SPT exhibits a higher sensitivity to small scales and
it does not detect the excess of lensing power at large multipoles,
the idea of replacing the Planck spectrum to that observed by the SPT
seems rather natural. Inspired by this simple idea, we present a combined
data approach which utilises both Planck and SPT measurements providing
consistent parameter inference from both large and small angular scales. For that, we restrict the multipole range of the
Planck TT spectrum to $\ell<1000$ and combine it with the SPTPol
measurements of TE and EE power spectra at $50<\ell\leq8000$. This allows us simultaneously to get rid of the lensing
tension, which affects the Planck spectrum at high multipoles, and
take benefit from the ground-based experiments where their sensitivity
surpass that of the Planck measurements. We proceed in similar fashion to Ref. \cite{Keisler:2011aw} that combines the CMB damping tail from the partial SPT-SZ survey and seven-year WMAP temperature power spectrum to improve constraints on cosmological parameters. We do not include the TT
spectrum at the high multipoles from SPT-SZ survey since this
measurement has lower statistical power as compared to the Planck TT
spectrum at $\ell<1000$ and does not influence our parameter
constraints.

An alternative avenue in handling the tensions between the low and high
redshift measurements is a modification of the cosmological concordance
model. Indeed, the parameter fit based on the CMB data is highly model
dependent and therefore influenced by possible extensions of
$\Lambda$CDM model. Many attempts have been made to restore
the concordance between different dataset by modifying either the early or
the local Universe physics. The late-time solutions include modified
\cite{DiValentino:2017zyq,Qing-Guo:2016ykt,Zhao:2017urm}, interacting
\cite{DiValentino:2016hlg,DiValentino:2017iww,DiValentino:2019ffd,DiValentino:2019jae,Kumar:2019wfs},
viscous \cite{Mostaghel:2016lcd,Wang:2017klo,Yang:2019qza} and
phenomenologically emergent dark energy models
\cite{Li:2019ypi,Pan:2019hac}, interaction between the dark sectors
\cite{Ko:2016uft,Raveri:2017jto,DiValentino:2017oaw,Archidiacono:2019wdp}
and decaying dark matter
\cite{Berezhiani:2015yta,Chudaykin:2016yfk,Poulin:2016nat,Chudaykin:2017ptd}. Recent
studies indicate that modification of the Universe expansion history in the two
decades of scale factor evolution prior to recombination is needed to
solve the $H_0$ problem \cite{Bernal:2016gxb,Aylor:2018drw}. The popular
early-time solutions are Early Dark Energy (EDE)
\cite{Karwal:2016vyq,Poulin:2018cxd,Agrawal:2019lmo,Lin:2019qug,Ye:2020btb}
and strong scattering interactions between the neutrinos or between
other additional light relics
\cite{Cyr-Racine:2013jua,Lancaster:2017ksf,Kreisch:2019yzn}. A
significant 
progress in clarifying different discrepancies has been made with
cosmology independent reconstruction technique of the Universe
evolution \cite{Joudaki:2017zhq,Poulin:2018zxs,Keeley:2019esp}. For 
the present status of the Hubble tension one can see \cite{Knox:2019rjx}.

In this paper, we examine possible tensions between the Planck
cosmology and astrophysical data with a suitable modification of the
early-time cosmology. As a reference, we consider the EDE model. 
The cornerstone of our analysis is the combined data approach. 
We discard the Planck temperature power spectrum at high multipoles and combine 
the remainder with polarization and lensing measurements from SPTPol survey.
This strategy eliminates the lensing tension
inherent in the Planck spectra on small scales and provides with
robust determination of the cosmological parameters. It also allows us to assess 
the impact of internal inconsistencies within the Planck data on cosmological tensions 
in the $\Lambda$CDM and EDE models.

Previous EDE analyses \cite{Poulin:2018cxd,Agrawal:2019lmo,Smith:2019ihp}
have shown this model can alleviate the Hubble tension, but the late-time amplitude of density fluctuation $\sigma_8$ increases as compared to $\Lambda$CDM, increasing tension with large-scale structure data. Ref. \cite{Hill:2020osr} greatly updates previous analyses by considering large-scale structure data in detail. They found that additional weak gravitational lensing and galaxy clustering data substantially weaken the evidence for EDE, as result of the tension between the larger values of $S_8$ needed
to fit the CMB and SH0ES data in the EDE scenario and the lower values of this parameter measured by large-scale structure surveys. In this work, we show that the combined data approach restores concordance amongst these measurements allowing for large $H_0$ values without substantially degrading the fit to large-scale structure data. For the first time, we derive EDE constraints from the Planck large-scale CMB anisotropy and the SPTPol data, which are consistent both with SH0ES data and weak lensing measurements along with other direct probes of clustering statistics.

The paper is organized as follows. In Section \ref{sec:stat} we
introduce all relevant datasets and describe our numerical procedure. In
Section \ref{sec:method} we give a detailed description of our
analysis method and present parameter constraints in $\Lambda$CDM
model. Section \ref{sec:ede} specifies the background and perturbed
dynamics of EDE model. In Section \ref{sec:ede2} we present our final
parameter constraints utilizing both the CMB and the low-redshifts
measurements within the EDE model. We discuss especially the importance of
the intermediate redshift data composed of the supernova dataset and BAO measurements for cosmological
inference. Finally, we conclude in Section \ref{sec:concl}.


\section{Data sets and numerical procedure}
\label{sec:stat}

\subsection{Cosmological data}
\label{sec:data}

In our analysis we exploit several different combinations of the
following datasets.

\begin{itemize}
    \item The Planck temperature \texttt{Plik} likelihood
          truncated at multipoles $30\leq\ell<1000$ and complemented with \texttt{Commander} power spectrum at low multipoles $\ell<30$ \cite{Aghanim:2018eyx}. We include all nuisance parameters and impose the same priors as in the Planck analysis \cite{Aghanim:2018eyx} for those. We refer to this likelihood as $\rm PlanckTT\text{-}low\ell$.
    \item The Planck EE likelihood at low multipoles $\ell<30$
          \cite{Aghanim:2018eyx}. Since the measurement of
          polarization anisotropies at large scales is essential to
          constrain the optical depth $\tau$ which breaks the harmful degeneracy between the amplitude of CMB spectra $A_s$ and $\tau$ we include this
          likelihood in all datasets.
    \item CMB polarization measurements from the 500$\deg$
          SPTPol survey which includes TE and EE spectra in the
          multipole range $50<\ell\leq8000$ \cite{Henning:2017nuy}. To
          obtain cosmological constraints we marginalise posteriors over six foreground parameters ($D^{\mathrm{PS_{EE}}}_{3000}$, $A_{80}^{TT}$, $A_{80}^{EE}$, $A_{80}^{TE}$, $\alpha_{TE}$, $\alpha_{EE}$), the super-sample lensing variance $\kappa$, two instrumental calibration terms ($T_\mathrm{cal}$, $P_\mathrm{cal}$) and two beam uncertainty shot noises ($A^1_{\mathrm{beam}}$,  $A^2_{\mathrm{beam}}$). We also impose flat priors on the first four of these and Gaussian priors on the rest in full compliance with \cite{Henning:2017nuy}. We apply appropriate window functions to transform theoretical spectra from unbinned to binned bandpower space. We denote this likelihood as $\rm SPTPol$.
    \item Measurements of the lensing potential power spectrum
          $C_\ell^{\phi\phi}$ in the multipole range $100<\ell<2000$
          from the 500$\deg$ SPTPol survey \cite{Wu:2019hek}. The
          lensing potential is reconstructed from a minimum-variance
          quadratic estimator that combines both the temperature and polarization CMB maps. To incorporate the effects of the survey geometry we convolve the theoretical prediction for $C_\ell^{\phi\phi}$ with appropriate window functions at each point in the parameter space. 
          We also perturbatively correct $C_\ell^{\phi\phi}$ for changes due to the difference between the recovered lensing spectrum from simulation and the input spectrum following Ref. \cite{Bianchini:2019vxp}.
          We refer to this measurement as Lens in our analysis.
    \item Combined weak lensing measurements of
          $S_8=0.762\pm0.024$ \cite{Joudaki:2019pmv} provided by a
          homogeneous analysis of KiDS, VIKING-450 and DES cosmic
          shear surveys. Strictly speaking, the full large-scale structure likelihoods can be approximated by a simple Gaussian prior only in the concordance $\Lambda$CDM model. However, the authors in \cite{Hill:2020osr} have justified this procedure for EDE models showing that the information content in large-scale structure data is almost enterily contained in the $S_8$ constraint. Guided by this observation, we impose the appropriate Gaussian prior on $S_8$ and refer to this measurement as $\rm S_8$.
    \item The Cepheid-based local measurement of the Hubble constant $H_0=74.03\pm1.42\kms$ \cite{Riess:2019cxk}. We implement the SH0ES measurement as the Gaussian prior on $H_0$ and call it as $\rm H_0$.
    \item Baryonic Acoustic Oscillation measurements based on the consensus BOSS DR12 analysis \cite{Alam:2016hwk} which combines galaxy samples at $z=0.38$, $0.51$, and $0.61$. We name this likelihood $\rm BAO$.
    \item Luminosity distances of supernovae Type Ia coming from
          the Joint Light-curve Analysis (JLA) using SNLS (Supernova Legacy Survey) and SDSS (Sloan Digital Sky Survey) catalogues \cite{Betoule:2014frx}. We denote this low-redshift probe as $\rm SN$.
    
\end{itemize}

As far as we implemented the SPTPol likelihoods within \texttt{Montepython} environment and the publicly release \footnote{\href{https://pole.uchicago.edu/public/data/henning17}{
            \textcolor{blue}{https://pole.uchicago.edu/public/data/henning17}}
    } \!\footnote{\href{https://pole.uchicago.edu/public/data/lensing19}{
        \textcolor{blue}{https://pole.uchicago.edu/public/data/lensing19}}
} was based on another MCMC sampler \texttt{CosmoMC} it makes sense to distribute our code among science community. Our likelihood code is publicly available at \href{https://github.com/ksardase/SPTPol-montepython}{https://github.com/ksardase/SPTPol-montepython} and can be used for various cosmological analyses.

\subsection{Numerical procedure}
\label{sec:MCMC}

All theoretical calculations are carried out in the publicly available
Boltzmann \texttt{CLASS} code \cite{Blas:2011rf}. We adopt spatially flat Universe and assume normal neutrino hierarchy pattern with the total active mass $\sum m_\nu=0.06\eV$. To investigate the
EDE model we modify background and perturbation equations in
\texttt{CLASS} in full compliance with Sec. \ref{sec:ede}. To recover
the posterior distributions in $\Lambda$CDM and EDE models we apply
the Markov chain Monte Carlo (MCMC) approach. For that we use the
publicly available MCMC code \texttt{Montepython}
\cite{Audren:2012wb,Brinckmann:2018cvx}. Marginalized posterior
densities, limits and contours are produced with the latest version of
the \texttt{getdist} package \cite{Lewis:2019xzd} \footnote{\href{https://github.com/cmbant/getdist}{
        \textcolor{blue}{https://github.com/cmbant/getdist}}
}. 

To calculate the effects of CMB lensing we use the Halofit suite
\cite{Smith:2002dz,Takahashi:2012em} which allows for modelling the
small-scale nonlinear matter power spectrum. These corrections are
negligible for the lensed CMB power spectra but become relevant for
the calculation of the lensing potential power spectrum 
$C_\ell^{\phi\phi}$ at high multipoles. We include the Halofit module in all the data procedures.


\section{Methodology and $\Lambda$CDM constraints}
\label{sec:method}

\begin{figure}
    \begin{center}
        \includegraphics[width=1.0\columnwidth]{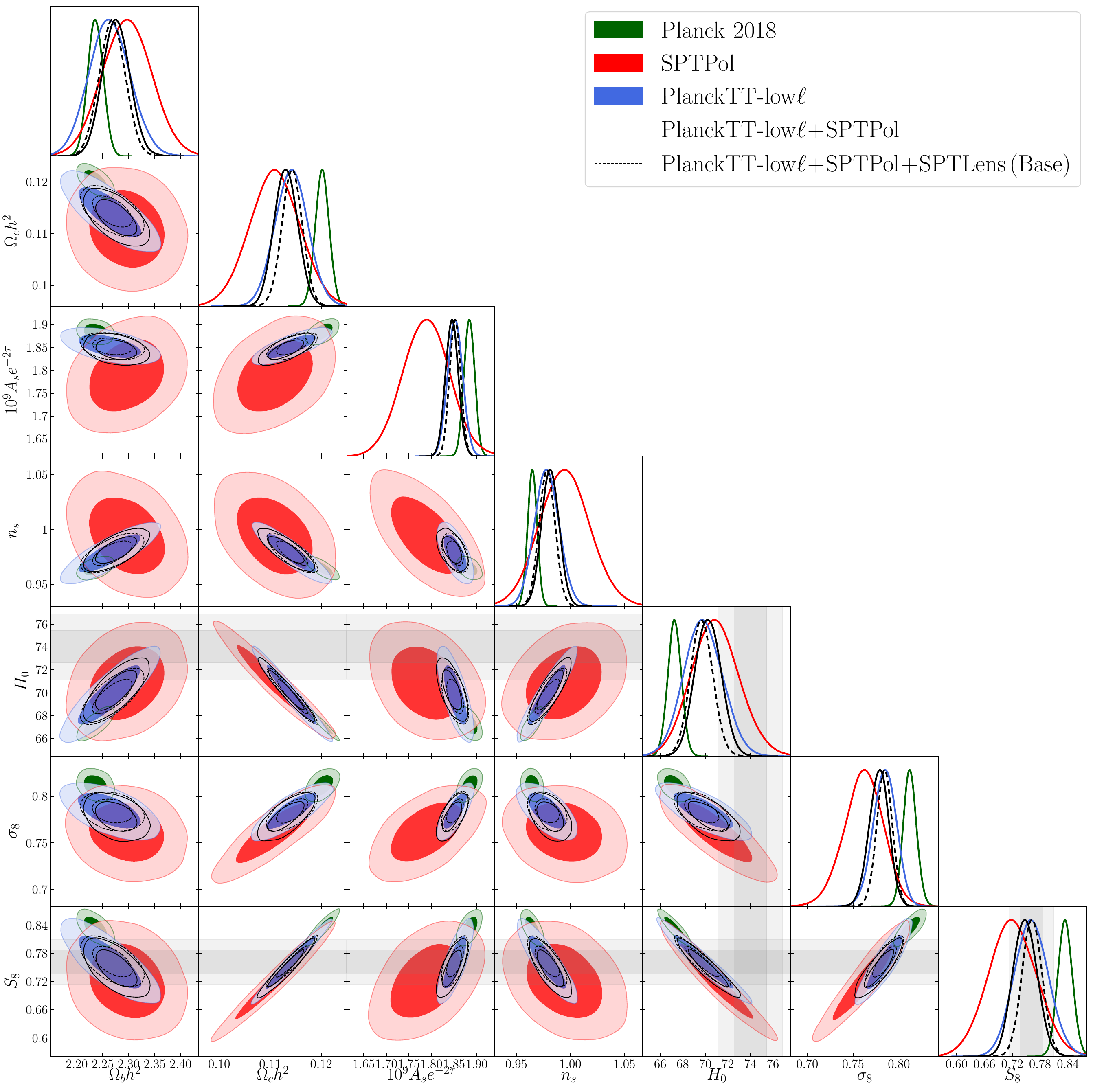}
        \caption {Marginalized parameter constraints for the $\Lambda$CDM model using different datasets. We explore independently $\rm PlanckTT\text{-}low\ell$ and $\rm SPTPol$, combined likelihood $\rm PlanckTT\text{-}low\ell\!+\!SPTPol$ along with SPTPol lensing, $\rm PlanckTT\text{-}low\ell\!+\!SPTPol\!+\!Lens$. For comparison we include constraints from the baseline Planck analysis $\rm PlanckTTTEEE$ as well. The gray bands represent the $1\sigma$ and $2\sigma$ constraints on $S_8$ and $H_0$ coming from \cite{Joudaki:2019pmv} and \cite{Riess:2019cxk}.}
        \label{fig:1}
    \end{center}
\end{figure}

As a preliminary step, we examine a consistency between $\rm
PlanckTT\text{-}low\ell$ and SPTPol datasets. For that we calculate
posterior distributions for relevant cosmological parameters which are
shown in Fig.~\ref{fig:1}.
We reveal that the parameter constraints inferred from $\rm
PlanckTT\text{-}low\ell$ (blue contours) and SPTPol (red contours)
likelihoods are perfectly consistent within $1\sigma$.  Thus, the two likelihoods are consistent with each other, and we proceed to combine them. To quantify the difference between our
approach and the standard data procedure we compare posterior
distributions for the combined likelihood $\rm
PlanckTT\text{-}low\ell\!+\!SPTPol$ (black contours) and the Planck
baseline analysis Planck 2018 (green contours). One may observe
that the resulting posteriors deviate from each other at
$\lesssim2\sigma$. In particular, our combined approach provides
substantially lower $S_8$ and higher $H_0$ as compared to the Planck baseline analysis, thus alleviating the tensions with local measurements. 

Parameter constraints inferred from $\rm PlanckTT\text{-}low\ell$, SPTPol and its combination are listed in Tab.~\ref{tab:1}. 
\begin{table}
    \renewcommand{\arraystretch}{1.1}
    \centering
    \begin{tabular} {| l | c |c | c | c |}
        \hline
        Parameter & $\rm \!PlanckTT\text{-}low\ell\!$ & $\rm SPTPol$  & $\rm \!PlanckTT\text{-}low\ell\!+\!SPTPol\!$ & $\rm Base$ \\
        \hline
        \hline
        $100\Omega_b h^2$ & $2.263\pm0.040$ & $2.296\pm0.048$ & $2.276\pm0.026$ & $2.269\pm0.025$  \\
        $\Omega_{c} h^2$ & $0.114\pm0.003$ & $0.111\pm0.005$ & $0.113\pm0.002$ & $0.114\pm0.002$ \\
        $H_0$ & $69.82\pm1.72$ & $70.90\pm2.12$ & $70.25\pm1.15$ & $69.68\pm1.00$ \\
        $\tau$ & $0.051\pm0.009$ & $0.053\pm0.010$ & $0.050\pm0.009$ & $0.051\pm0.009$ \\
        ${\rm ln}(10^{10} A_s)\!\!$ & $3.022\pm0.019$ & $2.989\pm0.031$ & $3.015\pm0.018$ & $3.021\pm0.017$ \\
        $n_s$ & $0.978\pm0.012$ & $0.995\pm0.023$ & $0.981\pm0.008$ & $0.979\pm0.007$ \\
        \hline   
        $\rdrag$ & $145.84\pm0.64$ & $146.47\pm1.34$ & $146.02\pm0.52$  & $145.76\pm0.46$ \\                                                           
        $\Omega_m$ & $0.283\pm0.020$ & $0.269\pm0.026$ & $0.277\pm0.013$ & $0.284\pm0.012$ \\
        $\sigma_8$ & $0.784\pm0.013$ & $0.762\pm0.021$ & $0.778\pm0.010$ & $0.784\pm0.009$ \\
        $S_8$ & $0.762\pm0.037$ & $0.721\pm0.052$ & $0.748\pm0.026$ & $0.763\pm0.022$ \\
        \hline
    \end{tabular}
    \caption {Parameter constraints in the standard $\Lambda$CDM
          model with $1\sigma$ errors. Polarization measurements at
          low multipoles are included in all datasets, see
          Sec.\,\ref{sec:data}. The Base dataset includes $\rm \!PlanckTT\text{-}low\ell\!+\!SPTPol\!+\!Lens$.}
    \label{tab:1}
\end{table}
We found that combining $\rm PlanckTT\text{-}low\ell$ and SPTPol
datasets one significantly improves constraints on all cosmological
parameters except for $\tau$ which mean value and errorbar remain 
intact. It happens because the corresponding constraint is driven
by the CMB polarization measurements on large scales alone which, in turn,
is included in all datasets. We stress that the value of $S_8$ is now
consistent with the homogeneous analysis of cosmic shear surveys
\cite{Joudaki:2019pmv} and with other direct probes of clustering
statistics \cite{Bocquet:2018ukq,Ade:2015fva,Vikhlinin:2008ym,Bohringer:2014ooa,Boehringer:2017wvr,Hildebrandt:2016iqg,Hildebrandt:2018yau,Abbott:2017wau,Abbott:2018ydy,Hikage:2018qbn}. The $\rm \!PlanckTT\text{-}low\ell\!+\!SPTPol\!$ dataset also
mitigates the Hubble tension to the $2.1\sigma$ level.

To break the degeneracy between the matter power spectrum $\sigma_8$
and the total matter density $\Omega_m$ we include measurements of
the lensing potential power spectrum. The Lens likelihood is primarily sensitive
to the parameter combination $\sigma_8\Omega_m^{0.25}$ which breaks the
degeneracy between $\sigma_8$ and $\Omega_m$ thus providing a more
robust constraint on the late time parameter $S_8$. The posterior
distribution for the $\rm \!PlanckTT\text{-}low\ell\!+\!SPTPol\!+\!Lens$
dataset is depicted in Fig.~\ref{fig:1} (dashed black lines) and
the corresponding parameter constraints are tabulated in
Tab.~\ref{tab:1}. We report that adding the Lens likelihood does not
induce any significant shifts in parameter constraints except for
$S_8$ which mean value increases by $0.6\sigma$ with somewhat reduced
error. It occurs because SPTPol TE,EE spectra dictate lower values of 
the inferred 4-point lensing amplitude with respect to SPTPol lensing measurements \cite{Bianchini:2019vxp} which results in higher $S_8$ upon adding the Lens likelihood.
Our result reveals the remarkable agreement between $\rm
\!PlanckTT\text{-}low\ell\!+\!SPTPol$ and $\rm Lens$ datasets that
justifies the using them in combination. In what follows we 
refer to the combined likelihood $\rm
\!PlanckTT\text{-}low\ell\!+\!SPTPol\!+\!Lens$ as {\it the Base set.} Final
constraints from the Base dataset read
\begin{equation*}
S_8=0.763\pm0.022\,, \qquad H_0=69.68\pm1.00\kms
\end{equation*}
We found that $S_8$ value is completely consistent with the local
probes, whereas the Hubble tension persists at the $2.5\sigma$ level.
We address this residual tension with one early-time solution in Sec.\,\ref{sec:ede2}.  

To justify statistical agreement amongst different likelihoods which constitute the Base dataset, we examine the goodness-of-fit to the CMB anisotropies as quantified by the $\chi^2$-statistic. In Tab.~\ref{tab:chi2_1} we list the $\chi^2$ values for each CMB likelihood for the best-fit $\Lambda$CDM model to the baseline Planck 2018 cosmology (second column) and Base dataset (third column).
\begin{table}
	\renewcommand{\arraystretch}{1.0}
	\centering
	\begin{tabular} {| c | c |c |}
		\hline
		Dataset & Planck 2018 best-fit & Base best-fit  \\
		\hline
		\hline
		$\rm {\it Planck}\, TT, \ell<30$ & $23.41$ & $20.93$ \\
		$\rm {\it Planck}\, EE, \ell<30$ & $396.19$ & $396.31$ \\
		$\rm {\it Planck}\, TT, 30\leq\ell<1000$ & $410.93$ & $404.54$ \\
		$\rm SPTPol$ & $150.04$ & $142.52$ \\
		$\rm SPTLens$ & $7.79$ & $5.21$ \\
		\hline
		Total $\chi^2$ & $988.36$ & $969.51$ \\
		\hline
	\end{tabular}
	\caption {$\chi^2$ values for the best-fit $\Lambda$CDM model to the baseline Planck 2018 cosmology and Base dataset.}
	\label{tab:chi2_1}
\end{table}
We found that the Base dataset significantly improves $\chi^2$-statistic for each likelihood except for the Planck EE $\ell<30$ data with respect to the Planck 2018 baseline analysis. It is caused by internal tensions within Planck data which results in different cosmological inference from low and high multipoles of Planck power spectra. The main culprit of this discrepancy is the overly enhanced lensing smoothing of the CMB peaks which pulls the late-time amplitude $\sigma_8$ to a higher value whilst the SPTPol measurements favour lower values of this parameter. Our data procedure is free from this tension and hence harvests a major improvement in the fit to both the Planck TT $\ell<1000$ and SPTPol likelihoods ($\Delta\chi_{\rm CMB}^2=-18.85$). Throwing away Planck data at high multipoles gives similar improvement \cite{Addison:2015wyg,Aghanim:2016sns,Aghanim:2018eyx}. Meanwhile, we found that the Base dataset notably worsens the $\chi^2$-statistic for SPTPol likelihood ($\Delta\chi_{\rm SPTPol}^2=3.93$) with respect to a pure SPTPol analysis which gives $\chi_{\rm SPTPol}^2=138.59$ \footnote{EE polarization measurements at large scales are also considered in this analysis.}.

\subsection{$A_L$ test}
\label{sec:Al} 

Here we present an important consistency check of cosmological constraints obtained using the combined data approach. Specifically, we verify that the Base dataset provides with 
a consistent lensing-induced smoothing of acoustic peaks within
the $\Lambda$CDM cosmology. For that, we introduce a free parameter $A_L$,
which scales the $C_\ell^{\phi\phi}$ at each point in the parameter
space. Thus scaled lensing potential power spectrum is used to lens
the CMB power spectra. Thereby, $A_L$ controls the theoretical prediction
for the lensing-induced smoothing effect. If the theory is correct and
the data is not affected by systematic effects, one expects $A_L=1$. In
what follows we examine $\Lambda$CDM model by making use of the combined data
approach.

Varying $A_L$ and 6 standard cosmological parameters of the $\Lambda$CDM concordance model we obtain almost identical posterior distributions with our baseline results outlined in Fig.~\ref{fig:1} (dashed black line) and listed in Tab.~\ref{tab:1} (Base dataset). We find 
\begin{equation}
A_L=0.990\pm0.035\,.
\end{equation} 
This result allows us to conclude that our combined data approach is
free from the lensing tension and provides consistent measurements of cosmological parameters in $\Lambda$CDM model. In general, combining the Planck TT power
spectrum in the multipole range $2<\ell<1000$ with the SPTPol measurements of
TE and EE spectra at $50<\ell\leq8000$ enables one to obtain an unbiased
cosmological inference from both large and small angular scales.

It is curious to examine the lensing-induced smoothing effect
in the $\rm PlanckTT\text{-}low\ell\!+\!SPTPol$ likelihood without 
Lens dataset. Following the same strategy we found in this case 
$A_L=0.935\pm0.050$. This constraint agrees with the less smoothing of 
acoustic peaks observed in SPTPol TE and EE spectra \cite{Henning:2017nuy}.
Our result reveals the importance of SPTPol lensing measurements which push the 
prediction into accordance with the $\Lambda$CDM model. 




\section{Early Dark Energy model}
\label{sec:ede}

To handle the residual tensions within $\Lambda$CDM model we resort to
an early-time solution. For the reference, we consider EDE that
behaves like a cosmological constant at early times and then dilutes
away with the Universe expansion like radiation or faster at later
times. If the dilution starts near the matter-radiation equality it results
in a larger Hubble constant and a smaller value of the baryon-photon
sound horizon thus alleviating the tensions with the direct cosmological
probes. In this Section, we state the homogeneous and perturbed
dynamics in the EDE sector pursuing generality and simplicity aims
for subsequent implementation in Sec.\,\ref{sec:ede2}.

First, we introduce one simple realization of EDE in the form of
the scalar field. Then, we consistently describe the homogeneous dynamics
of EDE field and evolution of its linear perturbations within an
effective-fluid approach \cite{Poulin:2018dzj}. We implement all
necessary equations in \texttt{CLASS} code and verify that they govern
correct background and perturbation dynamics. Finally, we obtain
actual constraints on both $\Lambda$CDM and EDE parameters, examine
various tensions and highlight the importance of $\rm BAO\!+\!SN$
dataset for cosmological inference.

It is worth noting that  the effective fluid description is applicable framework to track the scalar field dynamics as advocated in \cite{Smith:2019ihp}. The previous work \cite{Agrawal:2019lmo} claimed the wrong conclusion about the validity of using the approximate fluid approach. Why that study could not fully recover the results of Ref. \cite{Poulin:2018cxd} is easily explained by the different choice of the potential as shown in \cite{Smith:2019ihp}.

\subsection{Background dynamics}
\label{sec:background}

We consider EDE in the form of the cosmological scalar field $\phi$ with a power-law potential
\begin{equation}\label{V}
V_n(\phi)=V_0\frac{\phi^{2n}}{2^n}
\end{equation} 
where $V_0$ denotes the potential amplitude and $n$ states the
power-law index. At early times, the Hubble friction dominates and the
scalar field undergoes a "slow-roll" evolution. During this stage the
scalar field is frozen at its initial value $\phi_i$ acting as a pure
dark energy with equation of state $\omega_e\simeq-1$. Once the Hubble
parameter drops below a critical value $m$ (which is determined by a form
of potential through $m^2\simeq \d^2V_{n}/\d\phi^2$), the field starts to
oscillate around minima of its potential. During the oscillating
period the field amplitude decreases in time which dilutes
the field energy density. For $n=1$ the field undergoes simple
harmonic oscillations with a frequency which is independent of its
amplitude and for $n>1$ the oscillations are anharmonic and the 
frequency depends on the amplitude.  Rapidly oscillating solutions can
be modeled by an effective, averaged-over-cycle description with the effective equation
of state $\omega_e\simeq\omega_n$ given by
\begin{equation}\label{omega_n}
\omega_n=\frac{n-1}{n+1}\,.
\end{equation} 

\begin{figure}
    \begin{center}
        \includegraphics[width=0.48\columnwidth]{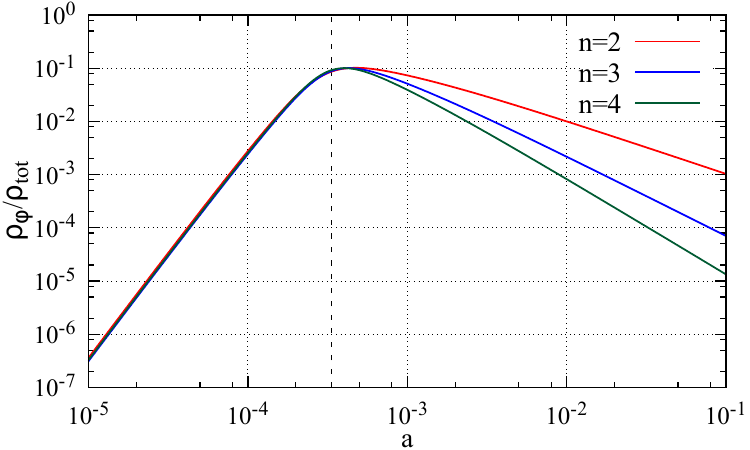}
        \includegraphics[width=0.48\columnwidth]{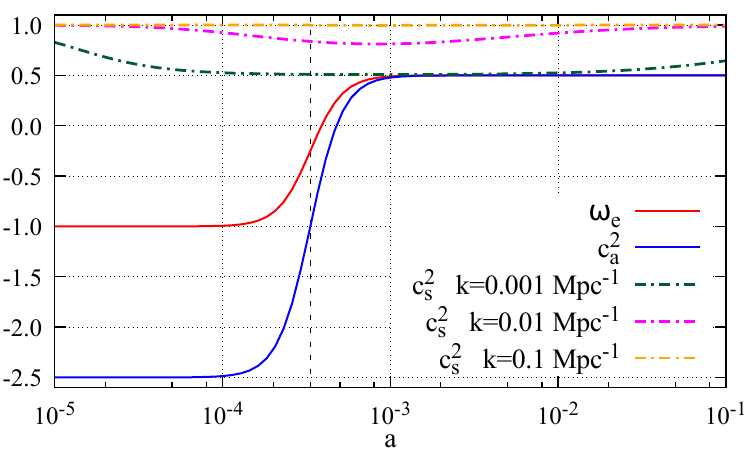}
        \caption { {\it Left panel:} evolution of the energy
                  density of the scalar field relative to total energy
                  density of the universe for several values of $n$
                  \ref{V}. {\it Right panel:} The scalar field
                  equation of state, the adiabatic sound speed $c_a^2$
                  and the effective sound speed $c_s^2$ for various
                  $k$ as a functions of scale factor. {\it Both
                    panels:} evolution of all quantities is obtained
                  with $f_e=0.1$, $z_c=3000$. The vertical dashed
                  black line refers to $z_c=3000$.}
        \label{fig:2}
    \end{center}
\end{figure}

To provide a smooth transition between the slow-roll and the oscillatory
periods we parameterise the energy density of the EDE field averaged
over the oscillation period by
\begin{equation}\label{rho_e}
\rho_e(a)=\frac{2\rho_e(a_c)}{1+\l a/a_c\r^{3(\omega_n+1)}}
\end{equation}
where $a_c$ refers to a transition between the two regimes. An
associated effective equation of state for the EDE fluid reads
\begin{equation}\label{omega_e}
\omega_e(a)=\frac{1+\omega_n}{1+\l a_c/a\r^{3(\omega_n+1)}}-1
\end{equation}
which asymptotically behaves as $\omega_e(a)\rightarrow-1$ at 
$a\rightarrow0$ and $\omega_e(a)\rightarrow \omega_n$ \eqref{omega_n}
for $a\gg a_c$. In Fig.~\ref{fig:2} we show the evolution of EDE
fraction in the total energy density of the Universe averaged over the
scalar field oscillations \eqref{rho_e} for several values of $n$ \eqref{V}
and the effective equation of state of EDE fluid \eqref{omega_e}.
On can see that the scalar field contribution to the total energy
budget of the Universe dilutes faster for larger $n$.

Ultimately, the background evolution of the EDE field is described by
parameters $\rho_e(a_c)$ and $a_c$. In what follows it is convenient
to use the maximum fraction of the total energy density in this field $f_e$
estimated over the all evolution history, $f_e$, 
\begin{equation}\label{f_e}
f_e\equiv\max\left\{\frac{\rho_e}{\rho_{\rm tot}}\right\}_{\text{all}\,a}
\end{equation}
We note that $f_e$ does not necessarily coincide with $\rho_e(a_c)/\rho_{\rm tot}(a_c)$ used in \eqref{rho_e}.

\subsection{Perturbed dynamics}
\label{sec:perturbed}

The effective fluid approach is a powerful tool to describe not only
a background evolution but also a perturbed dynamics of the rapidly
oscillating scalar field. This formalism provides with a set of
approximate 'averaged-over-cycle' equations in interms of fluid
variables. In order to utilize the approximate perturbation equations
of motion one needs the adiabatic sound speed $c_a^2$ and the effective
sound speed $c_s^2$. The adiabatic sound speed can be
straightforwardly calculated because it depends only on background
quantities
\begin{equation}\label{ca2}
c_a^2\equiv\frac{\dot{P}_e}{\dot{\rho}_e}=\omega_e-\frac{\dot{\omega}_e}{3(1+\omega_e)\mathcal{H}}\,,
\end{equation}
where $\mathcal{H}\equiv\dot{a}/a$ and the dot refers to the derivative with
respect to the conformal time. In Fig.~\ref{fig:2} we depict an evolution of
the adiabatic sound speed \eqref{ca2} as a function of the scale
factor. During the slow-roll evolution of the scalar field using
\eqref{omega_e} the adiabatic sound speed equals 
\begin{equation}\label{ca2_n}
c_a^2=-\frac{3n+1}{n+1}\,.
\end{equation}
Once the scalar field starts to oscillate, the adiabatic sound speed
reaches $\omega_e$ \eqref{omega_e}.

In order to compose the approximate perturbation equations we also
need the time-averaged effective speed sound in the fluid's rest
frame. We adopt the result from \cite{Poulin:2018dzj}
\begin{equation}\label{cs2}
c_s^2\equiv\Big\langle\!\!\Big\langle\frac{\delta P_e}{\delta\rho_e}\Big\rangle\!\!\Big\rangle=\frac{2a^2(n-1)\varpi^2(a)+k^2}{2a^2(n+1)\varpi^2(a)+k^2}\,,
\end{equation}
where $k$ denotes the conformal momentum and $\varpi$ is the
instantaneous oscillation frequency of the field fluctuations, which
for a pure power law potential is given by \cite{Poulin:2018dzj}
\begin{equation}
\varpi(a)\simeq \frac{3}{2}H(a_c)\sqrt{\frac{\pi}{n(2n-1)}}\frac{\Gamma(\frac{1+n}{2n})}{\Gamma(\frac{1+2n}{2n})}\frac{1}{\left[1+(a/a_c)^{3/(n+1)}\right]^{n-1}}\,.
\end{equation}
We note that $c_s^2=1$ for a slowly rolling scalar field but it
deviates from $1$ once the field starts to oscillate. This feature of
models with an early-time energy injection is crucial since $c_s^2<1$
over a large range of $k$ is needed to resolve the Hubble tension
according to Ref. \cite{Lin:2019qug}.

In the right panel of Fig.~\ref{fig:2} we plot the evolution of the effective sound
speed \eqref{cs2} with the scale factor for three values of $k$. It is
worth noting that the standard effective fluid approach assumes an abrupt
changing in the relevant quantities, specifically this implies $c_s^2=1$
and \eqref{ca2_n} at $a<a_c$, and \eqref{cs2} with \eqref{ca2} at
$a>a_c$. We verify that using the approximate, average-over-cycle quantities,
\eqref{cs2} and \eqref{ca2}, rather than the exact ones 
during all Universe evolution does not
lead to significant effects on the predicted power spectra used to
constrain the EDE model.

Finally, the equations which govern the evolution of density and
velocity perturbations in the synchronous gauge read
\begin{align}
\begin{split}\label{perturb}
&\dot{\delta}_e=-\left[u_e+(1+\omega_e)\frac{\dot{h}}{2}\right]-3(c_s^2-\omega_e)\mathcal{H}\delta_e-9(c_s^2-c_a^2)\mathcal{H}^2\frac{u_e}{k^2}\,,\\
&\dot{u}_e=-(1-3c_s^2)\mathcal{H}u_e+3\mathcal{H}(\omega_e-c_a^2)u_e+c_s^2k^2\delta_e\,,
\end{split}
\end{align}
where we introduce the heat-flux $u_e\equiv(1+\omega_e)\theta_e$ since
the linear perturbation equations written in terms of the bulk
velocity perturbation $\theta_e$ exhibit rapid growth being hard to track numerically. In principle, one should impose adiabatic initial conditions for $\delta_e$ and $u_e$ on super-Hubble scales. However, the EDE component is always subdominant on superhorizon scales at early times and its perturbations fall inside the gravitational potential created by the radiation component immensely fast. Given this reason, we take  $\delta_e=u_e=0$ initially since these quantities quickly approach a generic solution at radiation dominated stage \cite{Ballesteros:2010ks}.

\begin{figure}
    \begin{center}
        \includegraphics[width=0.48\columnwidth]{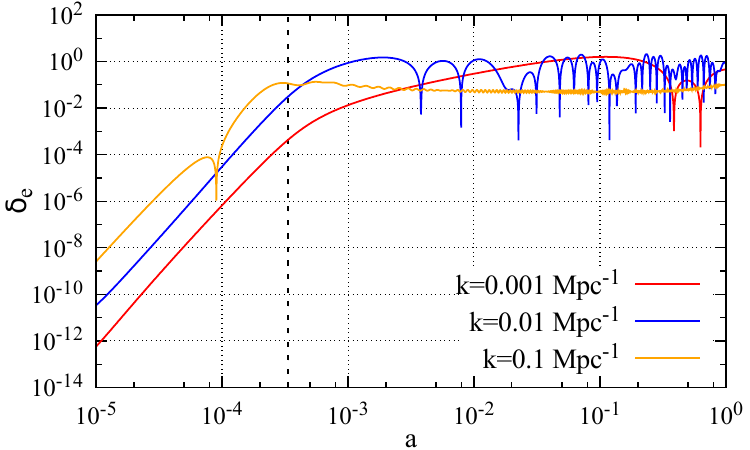}
        \includegraphics[width=0.48\columnwidth]{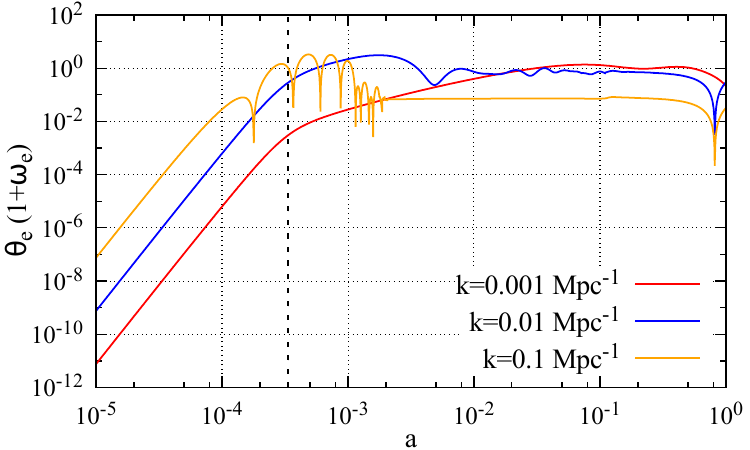}
        \caption { {\it Left panel:} the evolution of the
                  density contrast of the scalar field for a set of
                  momentum $k$. {\it Right panel:} the evolution of
                  the heat-flux for a set of $k$. {\it Both panels:}
                  the evolution of the quantities is obtained for
                  $f_e=0.1$, $z_c=3000$. The vertical dashed black
                  line refers to $z_c=3000$.}
        \label{fig:3}
    \end{center}
\end{figure}

In Fig.~\ref{fig:3} we depict the evolution of $\delta_e$ and $u_e$
with the scalar factor for a set of momentum $k$.  
As long as $c_s^2=1$, the pressure support leads to a strong decrease
in the perturbation amplitude for both superhorizon and subhorizon
modes. Once $c_s^2<1$, see \eqref{cs2}, the field internal pressure support decreases,
that  
yields nearly constant late-time density and velocity EDE
perturbations at $a>a_c$, seen in Fig.~\ref{fig:3}.



\section{Cosmological constraints within the Early Dark Energy model}
\label{sec:ede2}

EDE scenario can provide a larger Hubble constant and a smaller value
of the baryon-photon sound horizon. While the former trend presents an
opportunity to resolve the Hubble tension, the latter feature allows
one to reconcile the local measurements of cosmic distance-ladder with
the CMB observations. Indeed, $\rm H_0\!+\!BAO\!+\!SN$ brings a model-independent late-time estimate of the baryon-photon sound horizon at
the drag epoch $\rdrag$ (decoupling of protons from the cosmic
plasma), which is significantly lower than the
CMB-inferred value within the $\Lambda$CDM cosmology
\cite{Arendse:2019hev,Lyu:2020lwm}. This mismatch indicates a need for
modification of the early-time physics
\cite{Bernal:2016gxb,Aylor:2018drw}. In particular, reducing the
CMB-inferred sound horizon at the radiation drag epoch by $6-10\%$ would
reconcile the CMB-inferred constraints with the local $H_0$ and
$\rdrag$ determinations. It can be accomplished by an early energy
ejection prior to recombination
\cite{Karwal:2016vyq,Poulin:2018cxd}. In our study we examine this
possibility within the EDE framework using the combined data analysis
introduced in Sec.\,\ref{sec:method}.

In order to describe the background and perturbed dynamics of the EDE
field we modify the \texttt{CLASS} Boltzmann code implementing
\eqref{rho_e}, \eqref{omega_e} and \eqref{perturb}. We check that
these equations produce sensible physical outputs such as power
spectra of the EDE fluid. To derive the cosmological constraints within
the MCMC approach one needs to specify the parameter space of the EDE model.

The EDE model with a pure power-law potential \eqref{V} is fully
specified by three theory parameters: the redshift $z_c$ when the
scalar field starts to oscillate, the energy density of the EDE field
$\rho_e(z_c)$ at $z_c$ and the power-law index $n$ which parameterises the potential \eqref{V}. Due to large
scatter in $z_c$ and $\rho_e(z_c)$ variables we consider logarithmic
priors on these parameters \footnote{For the axion-like potential the uniform priors imposed on (physical) particle physics parameters (the axion decay constant and its mass) seriously downweight the preference for EDE models in comparison to uniform priors placed on the
	effective EDE parameters ($f_e$ and $\logg(z_c)$) \cite{Hill:2020osr}. In fact, this problem is attributed to the search of a proper theoretical solution that would match the prediction of our effective approach. Besides the axion-like potential which does not cover all physical realisations of the EDE scenario, there is a great variety of other EDE setups \cite{Ye:2020btb,Sakstein:2019fmf,Niedermann:2019olb} which have different physical priors. In our analysis, we hold a phenomenological point of view and vary the effective parameters, $f_e$, $\logg(z_c)$, which parameterise the EDE dynamics in a model independent  way.}. We note that absence of the initial scalar
field value $\phi_i$ in the definition of the effective sound speed
parameter \eqref{cs2} is caused by the power-law nature of our
potential \eqref{V}, see Ref.\,\cite{Smith:2019ihp}. Eventually, the
parameter space of the EDE model is characterized entirely by
$\logg(z_c)$, $\logg(\rho_e(z_c))$, $n$ and 6 standard parameters
$\omega_b$, $\omega_c$, $h$, $\tau$, $\ln(10^{10}A_s)$, $n_s$. In what
follows we consider $n$ as either fixed or free parameter in our
fitting procedure.

One comment is in order here. We emphasize that the fitting function calibration implemented in the Halofit module remains valid for EDE cosmology since the models capable of addressing the $H_0$ tension require $f_e\lesssim0.1$ which implies a small deviation from the $\Lambda$CDM expectation. More accurate justification for the validity of using the Halofit module is provided in \cite{Hill:2020osr}. Given this reason, we make use of the Halofit module in all EDE analyses.

\subsection{Parameter constraints for $n=3$}
\label{sec:const}

To impose robust constraints on the cosmological parameters, in what
follows we fix the shape of energy injection which is determined by parameter $n$, see \eqref{V}. The
data seem to favour lower values of the index $n$ since they provide a
larger peak energy injection fraction for a fixed width of the
transition from slow-roll to oscillatory regime resulting in a larger
Hubble constant and smaller value of the baryon-photon sound horizon
\cite{Smith:2019ihp,Agrawal:2019lmo}. On the other hand, lower $n$
models have slowly decaying with the scale factor tail as shown in
Fig.~\ref{fig:2}. Given this reason, it is difficult for these models
to inject sufficient energy in a relatively narrow time interval prior
to recombination while not having significant residual energy towards
low redshifts which, in turn, adversely affects the fit to the CMB
data. The fit to cosmological data is thus driven by the two competing
effects: injecting the largest possible amount of energy before
recombination and dilution of this energy as quickly as possible in
the post-recombination era. As a compromise between low and high
values of $n$ we choose $n=3$ in our baseline analysis; this choice
agrees with the recent EDE studies \cite{Smith:2019ihp,Hill:2020osr}. It is worth mentioning that we do
not consider $n=2$ which corresponds to the EDE diluting away like
radiation: it induces the phenomenon of self-resonance resulting in exponential growth of EDE perturbations \cite{Smith:2019ihp}. Since we
solve only linear perturbation equations for the EDE fluid, our
approximate framework can not address the resonance phenomenon
properly.

\begin{figure}
    \begin{center}
        \includegraphics[width=1.00\columnwidth]{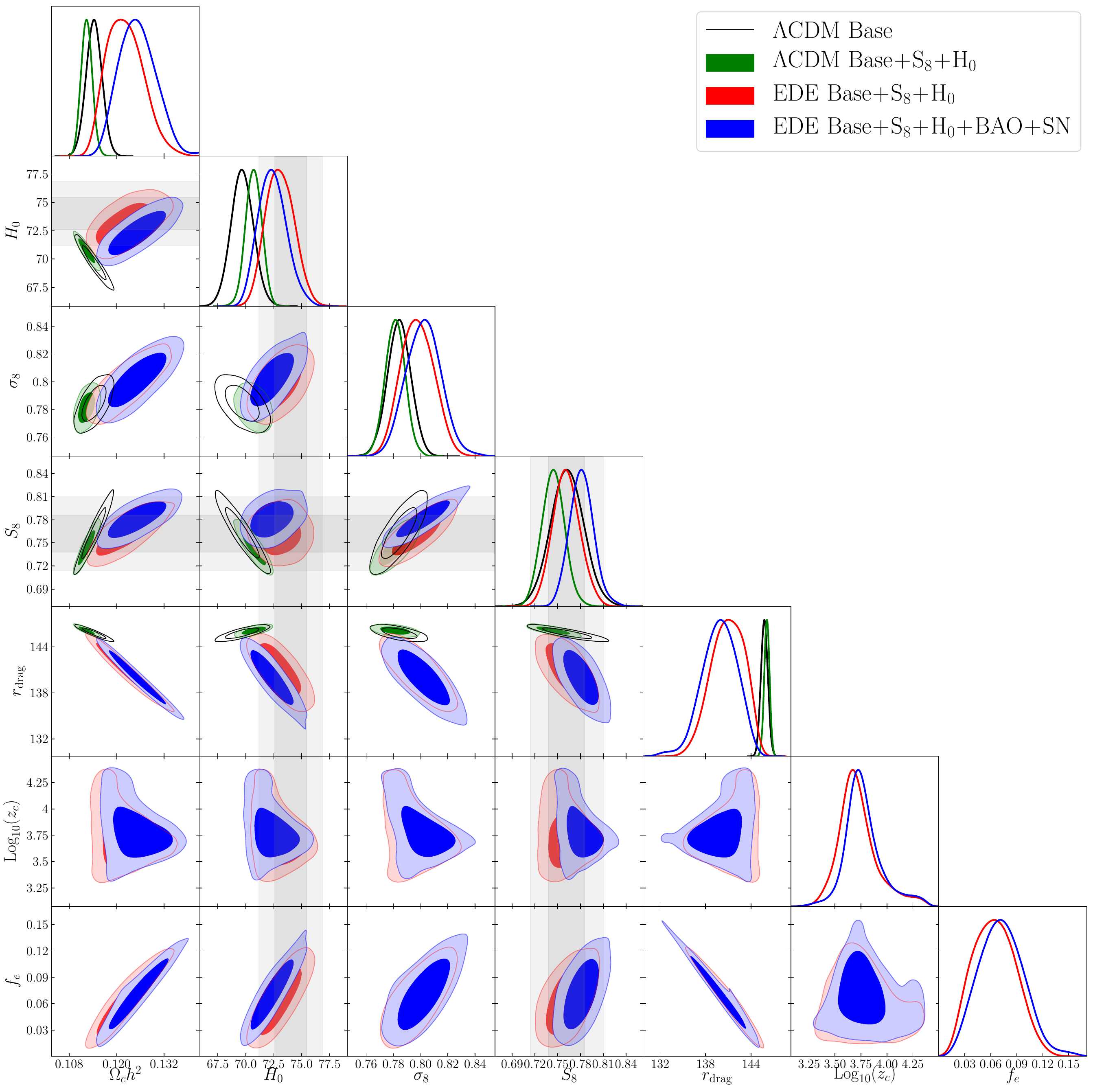}
        \caption {Marginalized parameter constraints for the
                  $\Lambda$CDM and EDE models using various 
                  datasets. We explore the 
                  parameter space using the
                  $\rm Base\!+\!S_8\!+\!H_0$ dataset with and without
                  intermediate redshift probe $\rm BAO\!+\!SN$. For
                  comparison we include constraints from the Base data
                  set only in $\Lambda$CDM model. The Base dataset
                  includes 
                  $\rm
                  \!PlanckTT\text{-}low\ell\!+\!SPTPol\!+\!Lens$. The gray bands represent the $1\sigma$ and $2\sigma$ constraints on $S_8$ and $H_0$ coming from \cite{Joudaki:2019pmv} and \cite{Riess:2019cxk}.}
        \label{fig:4}
    \end{center}
\end{figure}

The posterior distributions of the relevant cosmological parameters based on $\rm Base\!+\!S_8\!+\!H_0$ likelihood are shown in Fig.~\ref{fig:4} (red contours).
The corresponding parameter constraints are listed in Tab.~\ref{tab:2}. 
\begin{table}
    \renewcommand{\arraystretch}{1.1}
    \centering
    \begin{tabular} {| l | c |c | c |}
        \hline
        \multirow{2}{*}{Parameter} & $\rm \Lambda CDM$ & \multicolumn{2}{|c|}{${\rm EDE}\quad n=3$}    \\
        \cline{2-4}
        & $\rm Base\!+S_8\!+\!H_0\!$ & $\rm Base\!+\!S_8\!+\!H_0\!$  & $\rm Base\!+\!S_8\!+\!H_0\!+\!BAO\!+\!SN\!$  \\
        \hline
        $100\Omega_b h^2$ & $2.288\pm0.022$ & $2.314\pm0.037$ & $2.316\pm0.037$  \\
        $\Omega_{c} h^2$ & $0.112\pm0.001$ & $0.122\pm0.005$ & $0.126\pm0.005$ \\
        $H_0$ & $70.71\pm0.72$ & $73.06\pm1.26$ & $72.40\pm1.26$ \\
        $\tau$ & $0.056\pm0.008$ & $0.052\pm0.009$ & $0.048\pm0.009$  \\
        ${\rm ln}(10^{10} A_s)\!\!$ & $3.027\pm0.017$ & $3.032\pm0.017$ & $3.027\pm0.017$  \\
        $n_s$ & $0.985\pm0.006$ & $0.992\pm0.008$ & $0.987\pm0.008$  \\
        ${\rm Log}_{10}(z_c)$ & $-$ & $3.73\pm0.20$ & $3.78\pm0.19$  \\
        ${\rm Log}_{10}(\rho_e(z_c))$ & $-$ & $2.52\pm0.70$ & $2.76\pm0.67$  \\
        \hline
        $f_e$ & $-$ & $0.064\pm0.025$ & $0.072\pm0.026$  \\  
        $\rdrag$ & $146.10\pm0.35$ & $140.95\pm2.19$ & $139.86\pm2.35$      \\          
        $\Omega_m$ & $0.272\pm0.008$ & $0.274\pm0.008$ & $0.285\pm0.006$  \\
        $\sigma_8$ & $0.781\pm0.007$ & $0.798\pm0.012$ & $0.802\pm0.013$  \\
        $S_8$ & $0.744\pm0.015$ & $0.762\pm0.018$ & $0.782\pm0.016$  \\
        \hline
    \end{tabular}
    \caption {Parameter constraints in the standard $\Lambda$CDM
          and EDE models with $1\sigma$ errors. Polarization
          measurements at low multipoles are included to all datasets,
          see Sec.\,\ref{sec:data}. The Base dataset includes $\rm \!PlanckTT\text{-}low\ell\!+\!SPTPol\!+\!Lens$.}
    \label{tab:2}
\end{table}
We found that the Hubble tension is fully resolved in the EDE
scenario. Moreover, the fit does not degradate the $S_8$ constraint
remaining within $1\sigma$ interval with its local prediction
\cite{Joudaki:2019pmv}. Eventually, the dataset $\rm Base\!+\!S_8\!+\!H_0$
imposes $S_8=0.762\pm0.018$ and $H_0=73.06\pm1.26\kms$. Our outcomes are
qualitatively similar to results of the previous EDE analyses
\cite{Hill:2020osr,Smith:2019ihp,Agrawal:2019lmo} but, in contrast to them, we claim a perfect
consistency with the local measurements owing to the combined data
approach introduced in Sec.\,\ref{sec:method}.

A better fit to the local measurements is provided by the early-energy
injection just before recombination. Indeed, the marginalized
constraint on the redshift transition reads $\logg(z_c)=3.73\pm0.20$
which indicates the relatively brief energy injection before
recombination. In turn, the maximal injected EDE fraction equals
$f_e=0.064\pm0.025$. We emphasize that a quite brief period of the
energy injection around recombination minimizes the impact on other
successful $\Lambda$CDM predictions hence being essential for the good
fit to CMB. The most peculiar background feature in models with
early-time energy injection is a positive correlation between
$\omega_c$ and $f_e$ as plotted in Fig.~\ref{fig:4}. This behaviour
is attributed to the early integrated Sachs--Wolf effect which fixes
the height of the first CMB temperature acoustic peak. Indeed, the
extra energy injected by the oscillating scalar field can be
compensated by larger dark matter abundance thereby keeping the
Sachs--Wolf amplitude intact. At the level of perturbations, the
bigger dust component in the early Universe shifts the
matter-radiation equality to an earlier epoch. It elongates the matter
domination stage and affects the evolution of matter perturbations. In
particular, it increases a late-time amplitude of matter fluctuations
probed by $\sigma_8$. Thus, obtaining a bigger Hubble parameter and
a smaller sound horizon at the drag epoch within the models with energy
injection near recombination is accompanied by an increase of
$\sigma_8$ that is shown in Fig.~\ref{fig:4}. Furthermore, the
late-time parameter $S_8$ grows as well, that would help to distinguish EDE models from other competing solutions.

To justify the inclusion of Gaussian priors on $S_8$ and $H_0$ parameters we resort to statistical analysis. The $\chi^2$-statistic for each likelihood in the $\Lambda$CDM and EDE fits to the $\rm Base\!+\!S_8\!+\!H_0\!$ dataset is given in Tab.~\ref{tab:chi2_2}.
\begin{table}
	\renewcommand{\arraystretch}{1.0}
	\centering
	\begin{tabular} {| c | c |c |}
		\hline
		Dataset & $\Lambda$CDM & EDE  \\
		\hline
		\hline
		$\rm {\it Planck}\, TT, \ell<30$ & $20.22$ & $20.38$ \\
		$\rm {\it Planck}\, EE, \ell<30$ & $396.24$ & $395.98$ \\
		$\rm {\it Planck}\, TT, 30\leq\ell<1000$ & $407.00$ & $404.61$ \\
		$\rm SPTPol$ & $143.68$ & $142.14$ \\
		$\rm SPTLens$ & $5.67$ & $4.63$ \\
		$\rm S_8$ & $0.13$ & $0.29$ \\
		$\rm H_0$ & $6.58$ & $0.42$ \\
		\hline
		Total $\chi^2$ & $979.52$ & $968.45$ \\
		\hline
	\end{tabular}
	\caption {$\chi^2$ values for the best-fit $\Lambda$CDM and EDE models to the $\rm Base\!+\!S_8\!+\!H_0\!$ dataset.}
	\label{tab:chi2_2}
\end{table}
First, it is instructive to compare the $\Lambda$CDM fits to CMB measurements for the Base and $\rm Base\!+\!S_8\!+\!H_0\!$ datasets, the $\chi^2$ statistic for the former set is provided by Tab.~\ref{tab:chi2_1} (third column). We found that the goodness-of-fit to CMB data is moderately degraded upon imposing Gaussian priors on $S_8$ and $H_0$ ($\Delta\chi_{\rm CMB}^2=3.3$), predominantly driven by the worsened fit to the Planck TT $30\leq\ell<1000$ data. This change indicates that the distance-ladder Hubble measurement and CMB data are in tension within $\Lambda$CDM cosmology. Second, we confront the CMB fits to the $\rm Base\!+\!S_8\!+\!H_0\!$ likelihood in $\Lambda$CDM and EDE models. We found a significantly improved CMB fit in the EDE cosmology compared to $\Lambda$CDM ($\Delta\chi_{\rm CMB}^2=-5.07$), driven primarily by the restored concordance of the Planck TT $30\leq\ell<1000$ likelihood. Furthermore, the EDE model notably improves the $\chi^2$-statistic for CMB data using $\rm Base\!+\!S_8\!+\!H_0\!$ dataset with respect to the $\Lambda$CDM fit to the Base dataset ($\Delta\chi_{\rm CMB}^2=-1.77$). It implies that EDE restores concordance amongst CMB and SH0ES measurements providing even better fit to CMB data in comparison with the $\Lambda$CDM model without any priors on $S_8$ and $H_0$. We emphasize that the EDE model allows for larger $H_0$ values without substantially degrading the fit to the cosmic shear measurements. This result justifies the inclusion of additional large-scale structure data within EDE cosmology. It is instructive to compare our results with Ref. \cite{Hill:2020osr} that hints at the potential for additional large-scale structure likelihoods to substantially constrain the EDE models. The tight EDE constraints found there arise from the use of the full Planck likelihoods which pull the late-time amplitude $\sigma_8$ to higher values \cite{Addison:2015wyg,Aghanim:2016sns,Aghanim:2018eyx}. Thus, in order to simultaneously fit
the CMB and SH0ES data in the EDE model, one needs even higher values of the late-time amplitude, thereby conflicting with weak lensing and other large-scale structure probes. In contrast, when combining the Planck temperature power spectrum at low multipoles $\ell<1000$ and SPTPol data, we found substantially lower values of $\sigma_8$ allowing for the EDE solution of the Hubble tension being consistent with large-scale structure data.


An important cross-check of our parameter constraints can be provided by
intermediate redshift astrophysical data. The $\rm BAO\!+\!SN$ dataset
represents a late-time probe of the parameter combination $\rdrag
h$. An appealing characteristic of this measurement consists in its
direct nature since the supernova data allow for translating the
cosmic distance scale from the BAO observations to a given redshift in
a model-independent way. Then, one can calibrate the $\rdrag h$
measurement with the help of the local Hubble probe \cite{Riess:2019cxk}
to obtain a model-independent determination of the baryon-photon sound
horizon at the drag epoch $\rdrag$. On the other hand, the standard
ruler $\rdrag$ can be obtained through the CMB measurements but this
inference strongly depends on the cosmological model assumption. If
the late-time probe of $\rdrag$ given by $\rm BAO\!+\!SN\!+\!H_0$
matches its early-time estimate within the EDE scenario provided by the
CMB, it will be strong evidence in favour of the EDE model. We
examine this possibility by including the $\rm BAO\!+\!SN$ dataset in
what follows.

Resulting constraints from the $\rm
Base\!+\!S_8\!+\!H_0\!+\!BAO\!+\!SN$ dataset are shown in
Fig.~\ref{fig:4} (blue contours). We observe that the Hubble constant measurement is completely consistent with that from the $\rm Base\!+\!S_8\!+\!H_0$
dataset,
$H_0=72.40\pm1.26\kms$. On the contrary, the value of $S_8$ underwent
substantial raising by $1.1\sigma$ upwards to $S_8=0.782\pm0.016$. The
further increase of $S_8$ is disfavoured by the direct probes of
clustering statistics. It means that the upcoming galaxy and weak lensing
surveys will elucidate the potential of the EDE fluid to completely
reconcile the cosmological tensions.


\begin{table}[!t]
	\renewcommand{\arraystretch}{1.0}
	\centering
	\begin{tabular} {| c | c |c |}
		\hline
		Dataset & $\Lambda$CDM & EDE  \\
		\hline
		\hline
		$\rm {\it Planck}\, TT, \ell<30$ & $21.35$ & $20.49$ \\
		$\rm {\it Planck}\, EE, \ell<30$ & $395.66$ & $397.02$ \\
		$\rm {\it Planck}\, TT, 30\leq\ell<1000$ & $405.94$ & $405.45$ \\
		$\rm SPTPol$ & $142.90$ & $141.48$ \\
		$\rm SPTLens$ & $5.37$ & $4.47$ \\
		$\rm S_8$ & $0.01$ & $1.90$ \\
		$\rm H_0$ & $9.16$ & $0.50$ \\
		$\rm BAO$ & $7.06$ & $4.34$ \\
		$\rm SN$ & $683.38$ & $683.16$ \\
		\hline
		Total $\chi^2$ & $1670.83$ & $1658.81$ \\
		\hline
	\end{tabular}
	\caption {$\chi^2$ values for the best-fit $\Lambda$CDM and EDE models to the $\rm Base\!+\!S_8\!+\!H_0\!+\!BAO\!+\!SN$ data.}
	\label{tab:chi2_3}
\end{table}

The main advantage of the intermediate redshift astrophysical data is
that it yields the absolute scale for the distance measurements (anchor)
at the opposite end of the observable Universe. In particular, $\rm
H_0\!+\!BAO\!+\!SN$ provides with a late-time model-independent probe of
$\rdrag$ which value can be contrasted with the CMB inference under the assumption of the EDE model. We find that inclusion of $\rm
BAO\!+\!SN$ drives the sound horizon at the drag epoch downwards by
$0.5\sigma$ with the nearly identical error, $\rdrag=139.86\pm2.35$. This value
demonstrates a perfect consistency with the EDE prediction. Reducing
of $\rdrag$ with respect to the concordance $\Lambda$CDM model reflects the main
property of early-time energy injection models. For instance, the
sound horizon at the drag epoch in the EDE scenario is diminished in comparison with
the $\Lambda$CDM consideration by $4\%$ for both $\rm Base$ and $\rm Base\!+\!S_8\!+\!H_0$ datasets that agrees with the previous investigations
\cite{Bernal:2016gxb,Lyu:2020lwm} \footnote{It worth noting that our constraints on the sound horizon $\rdrag$ in the $\Lambda$CDM cosmology given in Tab.~\ref{tab:1} dictate substantially lower values of this parameter with respect to the Planck measurements \cite{Aghanim:2018eyx}. It is caused by higher values of $H_0$ dictated by our combined data approach which, in turn, results in lower values of $\rdrag$. Given this reason, the reduction in sound horizon at drag epoch by $4\%$ will suffice to reconcile the local Hubble measurements with CMB.}. Regarding the EDE sector, we obtain 
$f_e=0.072\pm0.026$ which indicates a $2.8\sigma$ evidence for nonzero
EDE fraction. 

To assess the concordance of $\rm BAO\!+\!SN$ data with the rest measurements, we provide the $\chi^2$-statistic for each likelihood in the $\Lambda$CDM and EDE fits to the $\rm Base\!+\!S_8\!+\!H_0\!+\!BAO\!+\!SN$ dataset in Tab.~\ref{tab:chi2_3}. The $\Lambda$CDM fit exhibits a slight tension with the BAO measurements at $1.5\sigma$ level \footnote{The value $\chi^2_{\rm BAO}=7.06$ is distributed as $\chi_N^2$ with effective degrees of freedom $N=4$ given by number of data points ($D_A$ and $H$ at three different redshifts) minus sum of fitting parameters which parametrize the theory prediction ($\omega_c$ and $h$).}. This mild tension is completely alleviated in the EDE scenario. The fit to supernova data is not worsened either in the EDE model compared to $\Lambda$CDM. Regarding the EDE fit to $\rm Base\!+\!S_8\!+\!H_0$ dataset given in Tab.~\ref{tab:chi2_2} (third column), we observe nearly the same contributions of all likelihoods to the $\chi^2$-statistic except for the cosmic shear measurements which notably degrade the EDE fit ($\Delta\chi_{\rm S_8}^2=1.61$) highlighting the specific role of future large scale structure data. These results emphasize the remarkable agreement between $\rm BAO\!+\!SN$ and $\rm Base\!+\!S_8\!+\!H_0$ likelihoods which justifies the combination of these measurements in one dataset.

\begin{figure}
    \begin{center}
        \includegraphics[width=0.48\columnwidth]{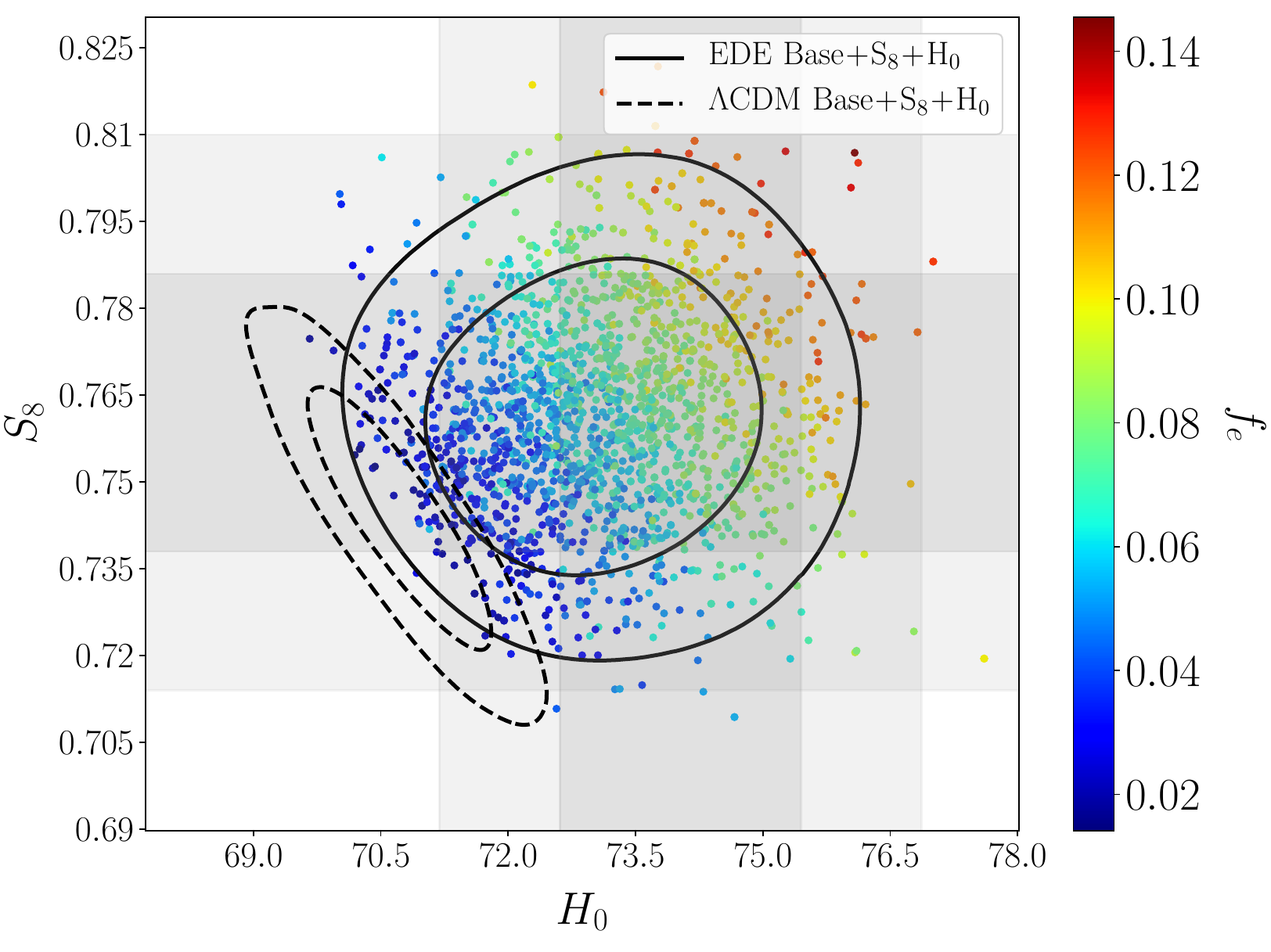}
        \includegraphics[width=0.48\columnwidth]{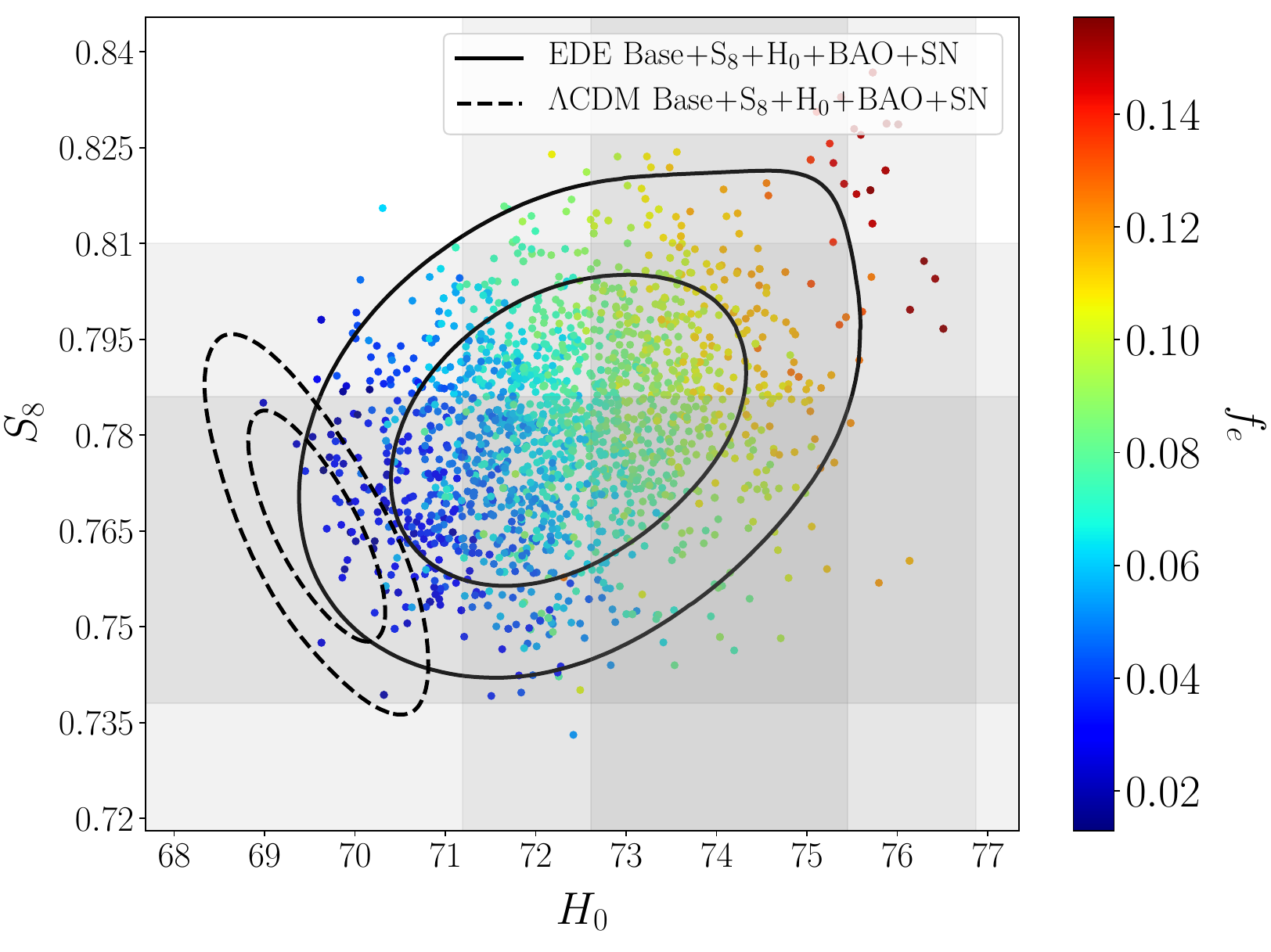}
        \caption {The marginalized posterior distribution in
                  the plane
                  $(S_8,\,H_0)$ for two data set combinations $\rm
                  Base\!+\!S_8\!+\!H_0$ ({\it left panel}) and $\rm
                  Base\!+\!S_8\!+\!H_0\!+\!BAO\!+\!SN$ ({\it right
                    panel}) in the $\Lambda$CDM and EDE models. The
                  scattered points represent values of $f_e$. The
                  gray bands represent the $1\sigma$ and $2\sigma$
                  constraints on $S_8$ and $H_0$ coming from
                  \cite{Joudaki:2019pmv} and
                  \cite{Riess:2019cxk}. The Base dataset includes $\rm
                  \!PlanckTT\text{-}low\ell\!+\!SPTPol\!+\!Lens$.}
        \label{fig:5}
    \end{center}
\end{figure}

Finally, it is instructive to examine the parameter constraints in the
plane $(S_8,\,H_0)$ with and without $\rm BAO\!+\!SN$ dataset. We show the
corresponding posterior distributions for EDE and $\Lambda$CDM models
in Fig.~\ref{fig:5}.
The negative correlation between $S_8$ and $H_0$ in $\Lambda$CDM model
can be readily understood. The parameter constraints in this case are
mostly driven by the CMB measurements which impose a tight constraint on the
observed angular size of the sound horizon at last scattering
$\theta_*$. For the base $\Lambda$CDM model, the main parameter
dependence of $\theta_*$ is approximately described by $\omega_c\cdot
h$ which should be constant. Given this reason, the higher Hubble constant
implies lower $\omega_c$. Since $\omega_c$ is positively correlated
with $\sigma_8$, the CMB anisotropy in the $\Lambda$CDM model yield
smaller values of $\Omega_m$ and $\sigma_8$ thus providing a lower
value of $S_8$ for higher $H_0$, see Fig.~\ref{fig:5} (dashed black
line). In the EDE case, the degeneracy direction in the plane $(S_8,\,H_0)$
is altered. The reason of that consists in different perturbation
dynamics in the EDE model witch results in amplified late-time
fluctuations of matter density probed by $\sigma_8$. It ensures a
negative correlation between $S_8$ and $H_0$ parameters shown in
Fig.~\ref{fig:5} (black line). Once we include the intermediate redshift
data $\rm BAO\!+\!SN$, the positive correlation between $S_8$ and $H_0$
parameters within the EDE scenario becomes more pronounced.

To understand quantitatively which model ($\Lambda$CDM or EDE) is
preferable, we corroborate our analysis based on posterior distributions with
$\chi^2$-analysis. For that, we compare the differences in logarithmic
likelihoods $\log L$ calculated for these two models in their
respective best-fit points for the same datasets. Each difference
$\Delta\log L$ is distributed as $\chi^2$ with effective degrees of
freedom equal to the difference in the number of free parameters in
$\Lambda$CDM and EDE models. 
Herein this number equals $2$ (recall in this Section we fix $n=3$)
corresponding to two extra parameters in
the EDE model, $\logg (\rho_e(z_c))$ and $\logg (z_c)$. Resulting 
improvements are shown in Tab.~\ref{tab:3}.
\begin{table}
    \begin{center}
        \begin{tabular}{|c|c|c|c|}
            \hline 
            Data set & ~$\Delta \chi^2$~ & ~p-value~ & Improvement\\
            \hline 
            $\rm Base\!+S_8\!+\!H_0$ & $11.1$ & $0.003887$ & $2.9\sigma$ \\
            $\rm Base\!+S_8\!+\!H_0\!+\!BAO\!+\!SN$ & $12.04$  & $0.00243$ & $3\sigma$ \\
            \hline
            $\rm Base\!+S_8\!+\!H_0\!+\!BAO\!+\!SN$ ($n$ free) & $12.04$ & $0.007247$ & $2.7\sigma$ \\
            \hline 
        \end{tabular}
    \end{center}
    \caption{A statistical improvement of EDE over $\Lambda$CDM
          in fitting the several data sets. In the top panel we set
          the power-law index to $n=3$ (2 extra degrees of freedom)
          wheres in the bottom line $n$ is treated as a free parameter
          (3 extra degrees of freedom).
        \label{tab:3}
    }
\end{table}

Our statistical analysis reveals that the EDE scenario strongly
improves the goodness-of-fit by $2.9\sigma$ compared to
$\Lambda$CDM. The preference does not change if one includes the
astrophysical data at intermediate redshifts $\rm BAO\!+\!SN$ that
indicates the remarkable agreement between the early and the late-time
cosmological inferences within the EDE model. We argue that the
early-time energy injection model provides a much better description
of the CMB and the low-redshift measurements of cosmological parameters.

\subsection{Promoting $n$ to a free parameter}
\label{sec:nfree}

So far, we have considered $n=3$, based on the cosmological
considerations.
However, from the phenomenological
point of view the shape of power-law potential \eqref{V} is not known
{\it a priory}. Hence, it is instructive to explore a
broader range of energy injection shapes which are controlled by
$n$. To this end we allow the power-law index $n$ to float freely in the
fit. To choose the appropriate range for $n$ we emphasize that for
a significantly large $n$ there is a specific class of no oscillatory (power law) solutions with an asymptotically constant equation of state \cite{Ratra:1987rm,Liddle:1998xm}. Since the resolution of the Hubble tension requires oscillatory solutions which ensure $c_s^2<1$ over a large range of $k$ \cite{Lin:2019qug}, we expect that the region of large $n$ is strongly disfavoured by the data \cite{Smith:2019ihp}. In our fitting procedure we impose the following flat prior on $n\in[1,6]$.

\begin{figure}
    \begin{center}
        \includegraphics[width=0.493\columnwidth]{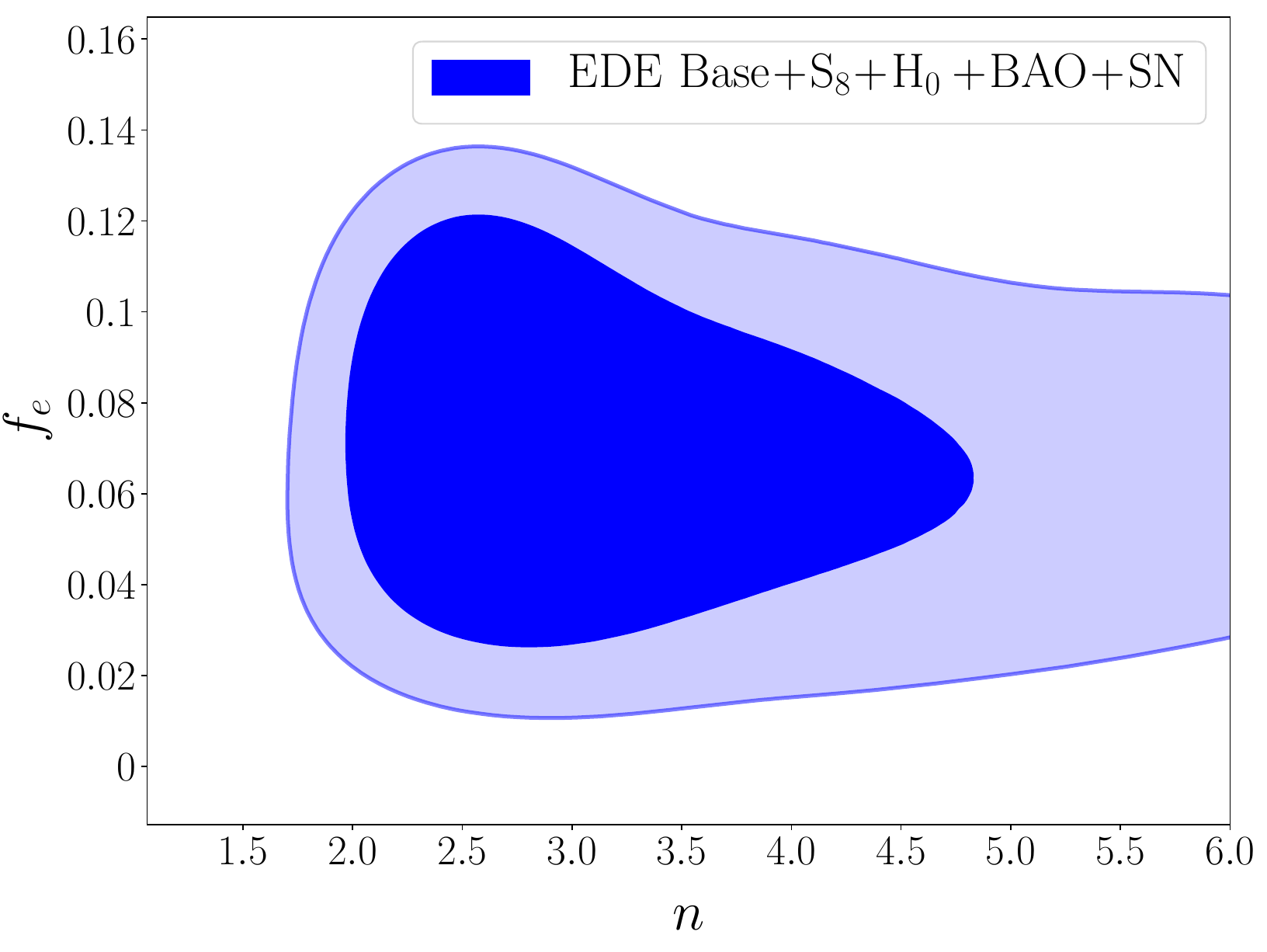}
        \includegraphics[width=0.48\columnwidth]{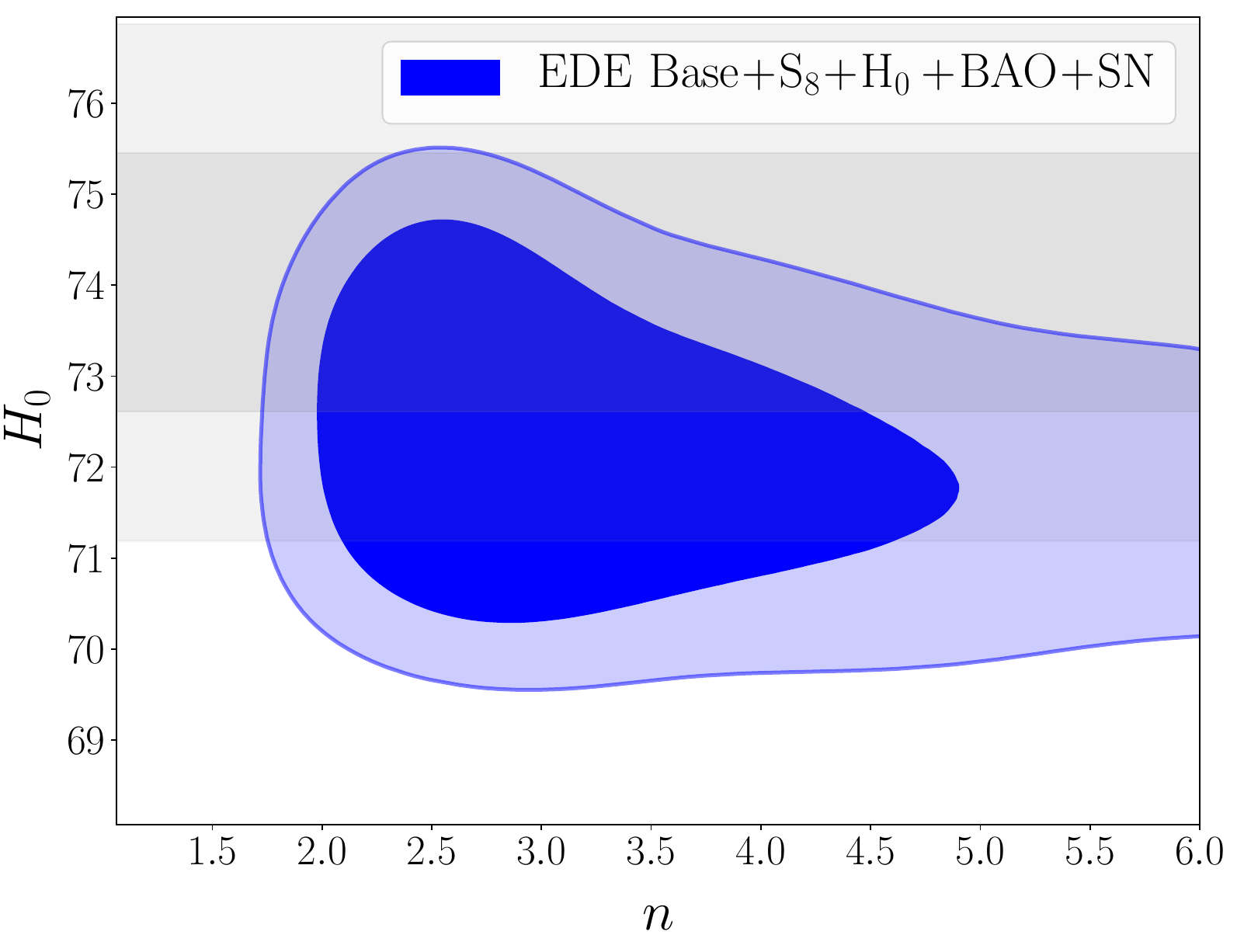}
        \includegraphics[width=0.7\columnwidth]{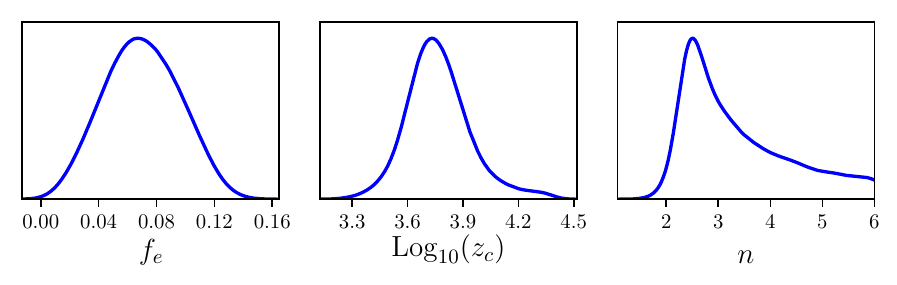}
        \caption {A marginalized posterior distribution in the
                  planes $(f_e,\,n)$ and $(H_0,\,n)$ using the $\rm
                  Base\!+\!S_8\!+\!H_0\!+\!BAO\!+\!SN$ dataset in the EDE
                  model. The gray bands represent the $1\sigma$ and
                  $2\sigma$ constraints on $H_0$ coming from
                  \cite{Riess:2019cxk}. The bottom panel represents 1d marginalized distributions for $f_e$, $\logg_{10}(z_c)$ and $n$ parameters. The
                  Base dataset includes $\rm \!PlanckTT\text{-}low\ell\!+\!SPTPol\!+\!Lens$.}
        \label{fig:6}
    \end{center}
\end{figure}

Our resulting posterior distributions are shown in Fig.~\ref{fig:6}.
As discussed above, the data tends to favor lower values of the index
$n$, since they allow for a larger peak energy injection fraction. On
the other hand, models with lower $n$ have slowly decaying tail which
adversely affect the fit to CMB data. Given this reason, the models
with $n<2$ dilute the injected energy too slow and therefore are
disfavoured, see the left panel of Fig.~\ref{fig:6}. The right panel
of Fig.~\ref{fig:6} clearly illustrates that larger values of the
Hubble constant are favoured for $2<n<3$. Our outcome is
qualitatively consistent with the previous EDE analyse
\cite{Agrawal:2019lmo} but extends this study to smaller values of
$n$. It has been established owing to effective fluid description
which allows for averaging the rapid field oscillations no matter how
high the rate is. An explicit numerical solution of the field
equations is very time consuming exercise that can make the oscillating field dynamics intractable. For instance, computational complexity restricts the previous analysis
performed in Ref.\,\cite{Agrawal:2019lmo} to the region $n>2$. It
explains why our parameter constraints are somewhat different at
$n\sim2$ as compared to those in \cite{Agrawal:2019lmo}.  It is worth
noting that the existence of the resonance at $n=2$ does not effect
our results since the resonance width is too narrow to be captured in
our analysis \cite{Smith:2019ihp}. We verified that the EDE
perturbations remain linear and never become comparable to the
homogeneous amplitude during the Universe evolution.


We carry out the $\chi^2$ statistical analysis for the case of varying
power-low index $n$. Our result is represented in Tab.~\ref{tab:3}. We do
not find any improvement with varying index $n$ compared to the case
$n=3$. Somewhat lowering of the overall improvement for the free $n$ analysis is caused by the penalty of adding one additional EDE
parameter, i.e. $n$. We confirm that the final result is almost
insensitive to the value of $n>2$ supporting the claim of
Ref.\,\cite{Poulin:2018cxd}.




\section{Conclusion}
\label{sec:concl}

We formulated a new method to analyse the CMB data, that combines the Planck
and the SPTPol data in a consistent way. This approach benefits from both
full-sky observations and ground-based experiments, and yields
an unbiased parameter inference. Using this approach we examine various
cosmological tensions in the $\Lambda$CDM and the early-time energy injection
model. We list our conclusions below.

\begin{itemize}
    \item $2.5\sigma$ tension between the CMB constraints and the cosmic
          shear measurements previously declared in
          Ref.\,\cite{Joudaki:2019pmv} is completely alleviated in
          the $\Lambda$CDM model. It also concerns other local probes of
          the late-time amplitude of matter density perturbation \cite{Bocquet:2018ukq,Ade:2015fva,Vikhlinin:2008ym,Bohringer:2014ooa,Boehringer:2017wvr,Hildebrandt:2016iqg,Hildebrandt:2018yau,Abbott:2017wau,Abbott:2018ydy,Hikage:2018qbn}. The upward
          shift of $\sigma_8$ inferred from the Planck analysis is solely driven by
          an excess of the lensing-induce smoothing of acoustic peaks in
          the Planck spectra which is absent in our approach. Our
          resulting constraint on the late-time parameter read
          $S_8=\sigma_8\sqrt{\Omega_m/0.3}=0.763\pm0.022$.
    \item Accounting for only the CMB measurements substantially diminishes the
          Hubble tension with local distance-ladder
          \cite{Riess:2019cxk} from $4.4\sigma$ to $2.5\sigma$ in
          the $\Lambda$CDM cosmology. The combined fit to the Planck 
          and the SPTPol
          data drives the $\Lambda$CDM fit to the remarkably higher
          value of the Hubble constant, $H_0=69.68\pm1.00\kms$.
    \item The residual tension with the local distance-ladder measurement of the Hubble
          constant \cite{Riess:2019cxk} is completely alleviated in the EDE
          scenario. Using $\rm Base\!+\!S_8\!+\!H_0$ dataset we find
          $H_0=73.06\pm1.26\kms$. At the same time, it does not degrade
          the fit to the direct probe of the late-time amplitude
          \cite{Joudaki:2019pmv} leading to
          $S_8=\sigma_8\sqrt{\Omega_m/0.3}=0.762\pm0.018$. It relieves 
          the conflict between the SH0ES-resolving EDE cosmologies and large-scale structure data recently claimed in Ref. \cite{Hill:2020osr}.
    \item The intermediate redshifts data $\rm BAO\!+\!SN$ present 
          important consistency check of the EDE model since it
          provides an 
          independent probe of $\rdrag h$ at late times. We found that
          the parameter constraints driven by $\rm H_0\!+\!BAO\!+\!SN$ are in excellent agreement with the EDE prediction. In particular, the $\rm
          Base\!+\!S_8\!+\!H_0\!+\!BAO\!+\!SN$ dataset yields somewhat lower value of the Hubble constant $H_0=72.40\pm1.26\kms$. At the
          same time, it results in a substantially higher amplitude of
          late-time matter perturbation characterized by $S_8=0.782\pm0.016$,
          but remaining within $1\sigma$ interval with its local
          measurement \cite{Joudaki:2019pmv}.
    \item The cosmological data disfavour the early-time
          injection models with the power-law indexes $n<2$. Furthermore,
          setting free the power-law index $n$ we do not find any improvement in comparison with the case of $n=3$.
    
    
\end{itemize} 

We emphasize that ongoing and future galaxy probes such as DESI,
Euclid and LSST will clarify the potential of early-time energy
injection model to completely alleviate various cosmological tensions.



\section*{Acknowledgments}

We thank Mikhail Ivanov for helpful discussions. We also thank Vivian Poulin for valuable comments. The work was supported by the RSF grant 17-12-01547. All numerical
calculations were performed on the Computational Cluster of the
Theoretical division of INR RAS and the MVS-10P supercomputer of the Joint Supercomputer Center of the Russian Academy of Sciences (JSCC RAS).

\appendix


\bibliographystyle{JHEP}
\bibliography{short}

\end{document}